\documentclass[aps,prb,twocolumn,notitlepage,showpacs,superscriptaddress]{revtex4-2}
\usepackage{graphicx}
\usepackage{amsmath,amssymb}
\usepackage[T1]{fontenc}
\usepackage{lmodern}
\usepackage[utf8]{inputenc}
\usepackage{natbib}
\usepackage{hyperref}
\hypersetup{colorlinks=true,citecolor={blue},linkcolor={blue},urlcolor={blue}}
\usepackage{units}
\usepackage[english]{babel}
\usepackage{cancel}
\usepackage[normalem]{ulem}
\usepackage{tcolorbox}
\usepackage{xcolor}
\usepackage{dsfont}
\usepackage{soul}

\begin{document}
\title{Critical states and anomalous wave transport in an aperiodic polariton monotile}

\author{Valtýr~Kári~Daníelsson}
\affiliation{Science Institute, University of Iceland, Dunhagi 3, Reykjavik, IS-107, Iceland}
\author{Helgi~Sigur{\dh}sson}
\email{helgi.sigurdsson@fuw.edu.pl}
\affiliation{Institute of Experimental Physics, Faculty of Physics, University of Warsaw, ul. Pasteura 5, Warsaw, PL-02-093, Poland}

\begin{abstract}
Recently ``the Hat'' monotile was introduced into the family of aperiodic tilings and quasicrystals boasting physical properties lying at the boundary of ordered and disordered systems. Here we study the two-dimensional wave transport, transverse localization and scaling properties of the quantum modes in a Monotile quasilattice. Our system is based on reconfigurable optical lattices for cavity-polaritons which provide flexible means to study wavepacket dynamics, strong nonlinear phenomena, and power-driven condensation in this new type of an aperiodic tiling. We confirm the existence of localized and critical states in the Monotile through direct diagonalization of the Schrödinger equation. Scaling analysis on the moments of the wavefunction distribution reveals anomalous transport regimes of super-diffusive and near sub-diffusive polariton transport associated with the fractal structure of the Monotile Hilbert space. We propose a strategy using resonantly excited polariton fluids to verify our findings.
\end{abstract}

\maketitle

\section{Introduction}
Quasicrystals are physical systems that lie between ordered translationally periodic systems and disordered amorphous systems~\cite{Suck2002}. They possess no unit cell or translational symmetry, making common approaches based on Bloch's theorem and the concept of Brillouin zones inadequate, but can regardless tile the plane (in case of two-dimensions) with no holes or overlaps. Their aperiodic long range order, self-similar eigenmodes and fractal spectrum boasts peculiar structural properties~\cite{Steurer2004}, critical states~\cite{Hiramoto1988}, and anomalous transport~\cite{Roche_JMP1997, Vardeny2013}. This has driven their investigation across many platforms such as ultracold quantum gases~\cite{Viebahn_PRL2019, Sbroscia_PRL2020, Yu2024}, magnonics~\cite{Watanabe2021}, plasmonics~\cite{Verre2014, Tsesses2025}, photonics~\cite{Freedman2006, Levi2011, Vardeny2013, Xu2021, Wang2024, Shi2024}, lasers~\cite{Vitiello2014, Arjas2024}, and cavity polaritons~\cite{Hendrickson2008, Poddubny_PRB2009, Baboux_PRB2017, Sturges2019, Alyatkin2025}.

%However, wave transport from aperiodically arranged local emitters that act as nonlinear oscillators which can display spontaneous ordering and synchronicity is much less studied.

%\hs{[Tala aðeins um munin á aperiodic, fraktal, og kvasikrystalla. - HS]}

Recently, a solid-state platform was developed to simulate various artificial crystalline systems with all-optical reconfigurability based on ballistic nonequilibrium Bose-Einstein condensates of microcavity exciton-polaritons~\cite{Tosi2012a, Ohadi_PRX2016, Topfer2020}. Exciton-polaritons are quasiparticles appearing in the strong light-matter coupling regime where cavity photons and quantum well electron transitions become tightly interwoven to form new half-and-half states of matter and light~\cite{Deng_RMP2010}. Possessing extremely fast optical response times and light effective masses, strong interparticle interaction strengths (i.e. optical nonlinearities), and full optical write-in and read-out, polariton fluids have allowed researchers to explore phases of macroscopic nonequilibrium quantum matter, vorticity, spin transport, and topological phenomena~\cite{Boulier2020}. Today, large-scale networks of polariton condensates can be optically generated across various lattice geometries, both regular~\cite{Berloff2017, Alyatkin2021, Alyatkin2024} and aperiodic~\cite{Alyatkin2025, alyatkin2026}.

Motivated by the recent demonstration of macroscopic coherence in aperiodic polariton condensate networks~\cite{Alyatkin2025, alyatkin2026}, we investigate the single-particle physics and mean-field dynamics of polaritons in the recently discovered aperiodic tiling generated as ``the Hat'' (or {\it einstein}) monotile~\cite{Smith2024} (see Fig.~\ref{fig:monotile_scheme}). Such a tiling can be constructed from an underlying deltoidal trihexagonal tiling (also known as the {\it tetrille} lattice). It can also be seen as a decoration of a $C_6$ symmetric aperiodic key tiling.~\cite{Socolar_PRB2023} For simplicity, we will refer to the Hat tiling as just the ``Monotile''. 
%Interestingly, despite being aperiodic, the Monotile has 6-fold rotational symmetry like normal periodic crystals~\cite{Socolar_PRB2023} which sets it apart from other quasiperiodic or quasicrystalline structures that cannot possess two-, three-, four- or sixfold rotational symmetry. \vkd{Spurning með þessa málsgrein, því ég held að sexhyrningssamhverfa sé ekki sérstaklega merkilegur eiginleiki fyrir ólotubundna grind, t.d. https://arxiv.org/pdf/2211.00127 hinsvegar vekur það náttúrlega ekki grun um ólotubundinn eiginleika að krystall hafi sexhyrningssamhverfu.}
%projections of a subset of six-dimensional hypercubic lattice points onto the two-dimensional tiling plane
%
% regular þýðir að flísarnar séu eins reglulegir marghyrningar, semiregular/arkímedísk ef flísarnar eru reglulegir marghyrningar, ekki endilega eins, og allir hornpunktar eru af sömu gerð.
%For completeness, we also compare our findings to the regular square lattice~\cite{Berloff2017, Alyatkin2021} and the aperiodic Penrose tiling~\cite{Alyatkin2025} which were recently explored in optical polariton lattices. Our results underline the unique interference properties of the monotile leading to strong wavefunction localization and dynamics that ...

%\section{Transverse localization properties}
We start our analysis by exploring the Monotile spectrum and eigenstates structure in Section~\ref{sec.states}. We verify through direct diagonalization of the two-dimensional (2D) Schrödinger equation and scaling analysis of the eigenstates that the monotile hosts localized, critical, and extended states all together. Similar to disordered systems, the appearance of criticality in the eigenstates is tied to a multifractal structure in the Hilbert space. We then explore in Sections~\ref{sec.transport} and~\ref{sec:nonherm} through direct numerical integration of both the 2D Hermitian and non-Hermitian Schrödinger equation how this criticality affects the transport and transverse localization properties of polariton waves in the Monotile. We calculate the diffusion exponent and temporal correlation exponent of a spreading wavepacket which determine the power-law shape of the wavepacket at long times. We identify parameter regimes of the optical lattice displaying superdiffusive and near subdiffusive transport of polaritons. Last, in Section~\ref{sec.cond} we extend our results to a full nonlinear mean field model of polariton quantum fluids and propose strategies to verify our findings.

\section{Localized, Critical, and Extended States} \label{sec.states}
\begin{figure}[t]
    \centering
    \includegraphics[width=\linewidth]{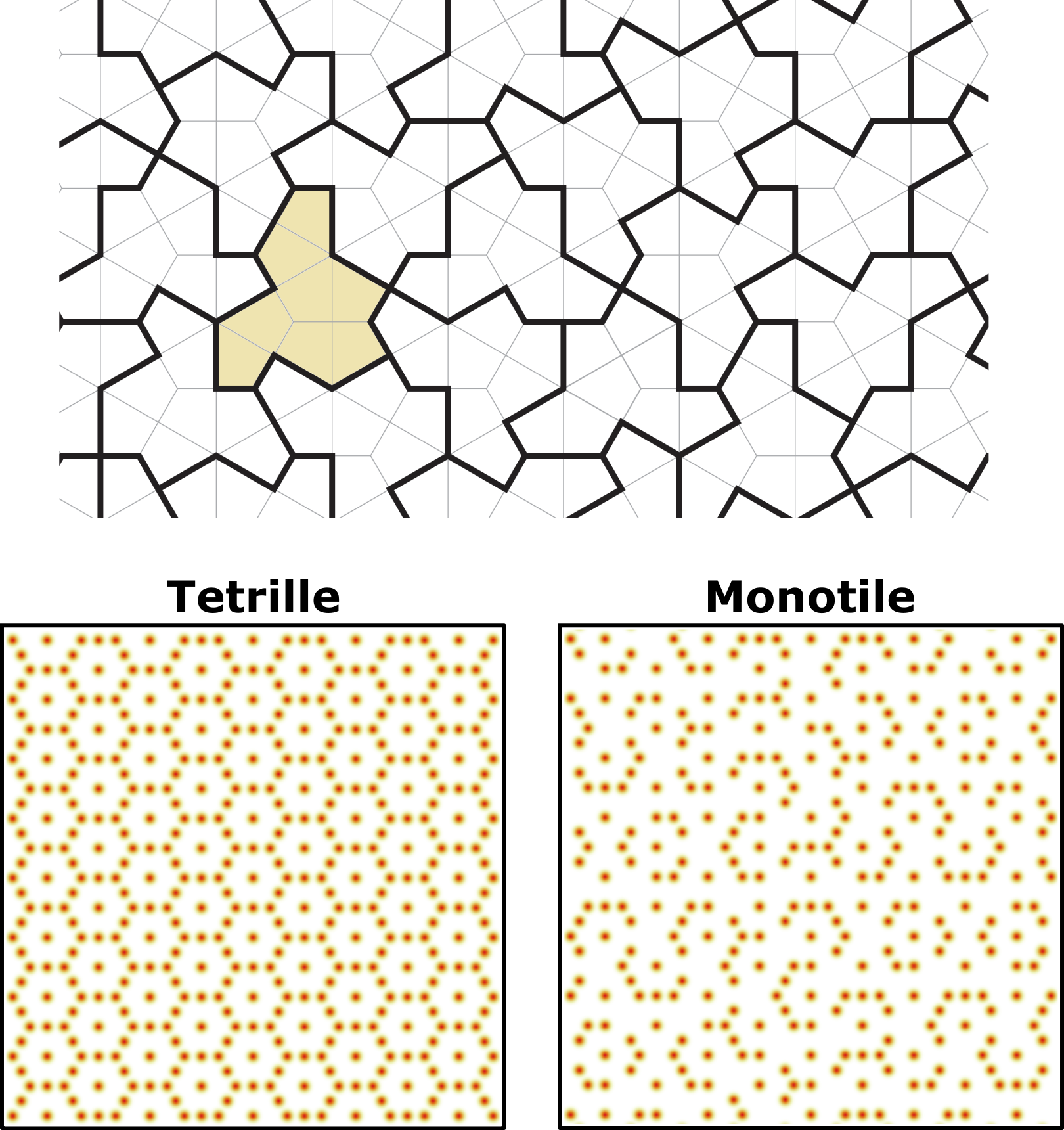}
    \caption{(Top) A patch of ``the Hat'' Tile($1,\sqrt{3}$) aperiodic monotile (thick edges) drawn on top of the [3.4.6.4] Laves tiling made from 6 equal kites in a hexagon illustrated with thin edges~\cite{Smith2024}. (Bottom) An example optically induced polariton potential $V(x,y)/V_0$ for the tetrille motherlattice and the corresponding Monotile quasilattice. Potential maxima (red spots) are located at the vertices of the geometries.}
    \label{fig:monotile_scheme}
\end{figure}
% Væri mögulega betra að byrja kaflann með að fjalla um eigingildisverkefnið og reikna meðal-IPR fyrir öll eiginástöndin upp að ákveðnri orku (t.d. 2V_0). 

%We start our analysis by exploring the transverse localization properties of polaritons in the four types of optical lattices. 

At sufficiently long in-plane wavelengths lower branch exciton-polaritons in a planar microcavity are accurately described by the 2D Schrödinger equation~\cite{Deng_RMP2010},
\begin{equation} \label{eq.schro}
    i \hbar \frac{\partial \psi}{\partial t} = \left[- \frac{\hbar^2 \nabla^2}{2m} + V(x,y) \right] \psi
\end{equation}
where $m$ is the effective polariton mass, $\nabla^2 = \partial_x^2 + \partial_y^2$, and $V(x,y)$ is the optically induced potential. For the moment we will neglect any cavity losses and amplification coming from the pump (i.e. $\text{Im}{(V)} = 0$) and study the Hermitian problem. This allows us to draw a comparison with other physical platforms used to study aperiodic systems but do not suffer from finite particle lifetimes. We will visit the non-Hermitian problem in section~\ref{sec:nonherm} focused on wave transport and condensation dynamics of polaritons in the Monotile in section~\ref{sec.cond}. 
%
% \begin{figure*}
%     \centering
%     \includegraphics[width=0.9\linewidth]{fig_pumps.png}
%     \caption{Pump induced potential $V(x,y)/V_0$ of three investigated geometries: Monotile, tetrille, and the P3 Penrose. %Here the nearest neighbour distance is $10$ $\mu$m.
%     }
%     \label{fig.pump_scheme}
% \end{figure*}

The optically induced potential originates from a photoexcited background density of excitons which repulsively interact with the polaritons. The much larger effective mass of the exciton makes the potentials appear static to polaritons. This allows all-optical tailoring of potential landscapes by simply adjusting the transverse profile of a nonresonant continuous wave beam incident upon the cavity. This also means that many types of potentials can be constructed, erased and rewritten on the same cavity sample~\cite{Berloff2017, Alyatkin2021}. We consider the case of a spatially uniform cavity (negligible disorder) and a potential constructed from identical Gaussian pump spots,
\begin{equation} \label{eq.V}
    V(x,y) = V_0 \sum_{n=1}^N e^{-|\mathbf{r}-\mathbf{r}_n|^2 / 2 w^2}.
\end{equation}
Here, $w$ is the root-mean-square width of the pumps and $V_0>0$ is the potential amplitude (proportional to the pump intensity and the exciton two-body interaction strength). We consider a vertex-pump geometry where each pump is located at the vertices of the corresponding lattice or tiling. We stress that this makes our system quite different from conventional photonic and condensed matter systems where traps (attractive potentials) are usually placed at the nodes (vertices) of the given geometry. Instead, the vertices in the optical potential form repulsive scatterers that diffract propagating polariton waves in the open space between them~\cite{Alyatkin2021}. Figure~\ref{fig:monotile_scheme} shows example pumping geometries $V(x,y)$ for the Monotile and its corresponding tetrille motherlattice. Notice how the Monotile is constructed from the tetrille with specific spots removed~\cite{Smith2024}. We have made our codes to generate the coordinates of the Tile($1,\sqrt{3}$) Monotile publicly available here~\cite{code}. The inverse problem of attractive potentials $V_0 \to -V_0$ can also be realized using e.g. coupled micropillars~\cite{Goblot2020} or cavity mesas but would be more technically challenging to fabricate and lack in-situ tunability. 
%We leave this problem for a future study.

%Since aperiodic systems don't possess the necessary symmetries to be described elegantly by Bloch states and a Brillouin zone, one must resort to alternative techniques to explore their properties. 
We apply direct diagonalization and scaling-analysis techniques, commonly used to study disordered quantum systems, and reveal the presence of critical states in the Monotile connected with a multifractal structure in the underlying Hilbert space. Criticality of the eigenstates is often found around phase transition points, such as the Anderson localization in the Aubry–André model~\cite{Goblot2020} or disordered materials~\cite{Evers_RevModPhys2008}. Aperiodic systems can also possess critical states in their spectrum without any such transition points~\cite{Saul1988}. Traditionally, 1D aperiodic systems have been used to study criticality and only recently have theoretical activities shifted more towards 2D aperiodic systems~\cite{Mace_PRB2017, Zhu_PRA2024, Duncan_PRB2024, Koga2024} but still mostly concerning vertex (tight-binding) models. We extend this effort towards the Monotile in a full 2D Schrödinger picture for a polariton scatterer potential.

Criticality manifests as self-similarity, nonergodicity, and a power-law spatial decay in the states of the system which give rise to exotic transport properties that are not fully ballistic or diffusive. The Hilbert space becomes multifractal with an infinite set of critical exponents $D_q$ that describe the scaling of the moments of the wavefunction probability distribution $P_q = |\psi(\mathbf{r})|^{2q}$~\cite{Evers_RevModPhys2008}. 
%Such critical behaviour is best known in metal-insulator transitions in disordered media but has also been observed in other complex objects such as turbulent and chaotic systems, and quasicrystals. 
The moments of the distribution are also known as the $q$th-order inverse participation ratios (IPR), 
\begin{equation}
    I_q  = \int |\psi|^{2q} d\mathbf{r},
\end{equation}
with normalization giving $I_1 = 1$. The IPR offers a useful measure on the amount of localization for a given state when no such prior knowledge exists. Large IPR values imply localized states whereas small IPR values imply extended (delocalized) states. 
%In particular, $P = 1/I$ is referred to as the {\em second participation moment} and has been commonly used as a measurement of the number of sites (in a lattice model) that participate in the wave function. 
A perfectly localized state gives $I_q = 1$ whereas a perfectly delocalized state is equally spread across the entire system with $I_q = 1/L^d$ where $L$ is the system side length and $d$ its dimension. The case $q=2$ is most often studied as it can be related to the effective (average) width of the wavefunction, $w_\text{eff} = 1/\sqrt{I_2}$. That is, how much area in the system ``participates'' in the wavefunction.

The IPR can reveal anomalous scaling with system size $L$ that distinguishes extended and localized states through a set of critical exponents or generalized (fractal) dimension $D_q$~\cite{Evers_RevModPhys2008},
\begin{equation} \label{eq.Dq}
    I_q   \sim L^{-D_q (q-1)}.
\end{equation}
Extended plane-wave like states have $D_q = d$ whereas for localized states one gets $D_q=0$. At criticality the state is neither extended or localized and the exponents  obtain a nontrivial dependence on $q$ in the range $0<D_q<d$ associated with a multifractal Hilbert space~\cite{Evers_RevModPhys2008}. In this context, the generalized dimension $D_2$, which measures the size of a wavefunction, is similar to the box-counting dimension in fractal analysis.%~\cite{Geisel_PRL1991}.
%In particular, if $D_q = D = \text{const.}$ then the system is monofractal whereas if $D_q$ depends on $q$ then it is multifractal. 
%
\begin{figure}
    \centering
    \includegraphics[width=\linewidth]{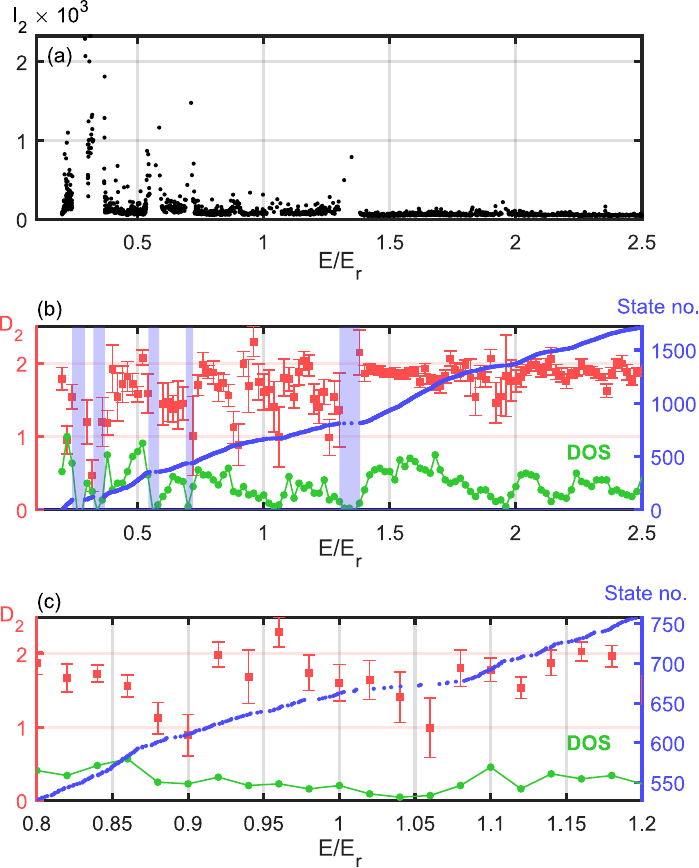}
    \caption{(a) IPR for the second moment $I_2$ of the Monotile eigenstates plotted against energy. (b) Double vertical-axis plot showing (left) the generalized dimension $D_2$ for the Monotile eigenstates as red squares and (right) corresponding eigenenergies $E_n/E_r$ as blue dots. The density of states (DOS) is additionally overlaid as green circles with an arbitrary vertical axis. Blue shaded regions mark the biggest pseudogaps in the spectrum. (c) Same as above but zoomed in on a smaller region of the spectrum.}
    \label{fig.D2_vs_En}
\end{figure}

We solve the eigenvalue problem $E\psi = \hat{H}\psi$ with open boundary conditions using a finite-difference method where $\hat{H}$ is the left-hand-side of the Schrödinger equation~\eqref{eq.schro}. The nearest neighbour spatial separation distance $a = 7$ $\mu$m is the characteristic distance of our calculations. Spatial resolution is set to $\Delta_x = a/10$ and energy will sometimes be given in units of the ``recoil energy'' $E_r = \pi^2 \hbar^2/2ma^2$ for convenience. We choose a typical value of the polariton mass in planar cavities $m = 5.6 \times 10^{-5} m_e$ where $m_e$ is the free electron rest mass and a potential amplitude $V_0 = 1 \text{ meV} \approx 7 E_r$ meV typical for the pump-induced blueshifts observed in experiment~\cite{Ohadi_PRX2016}. We solve the eigenvalue problem for different system sizes $L = 20a \to 40a$. To retain numerical accuracy, we only keep eigenstates up to $5 E_r$ in energy. Eigenstates that are localized close to the system boundary are discarded.  

Figure~\ref{fig.D2_vs_En}a shows the second moment $I_2$ of the Monotile eigenstates as a function of their eigenenergy. High $I_2$ values imply localization which mostly occurs at lower energies around the Monotile pseudogaps. Figure~\ref{fig.D2_vs_En}b shows the lowest sorted eigenenergies (blue dots) for a system size $L=40a$ and the corresponding generalized dimension $D_2$ in red and density of states in green (DOS). The largest pseudogaps found in the spectrum are marked with shaded blue regions. In fact, such pseudogaps exist at all scales due to the fractal nature of the spectrum.
%\hs{[Ræða meira um þetta. - HS]}. 
The generalized dimension was obtained by binning eigenstates of similar energy within a small window $[E-\delta,E+\delta]$ where $\delta = 0.01 E_r$ and performing simple linear regression of $I_2$ against $L$ on a log-log scale. One standard error is indicated by the red errorbars. 
%We additionally filtered out any states of unusually high IPR corresponding to localized states.

\begin{figure*}
    \centering
    \includegraphics[width=0.9\linewidth]{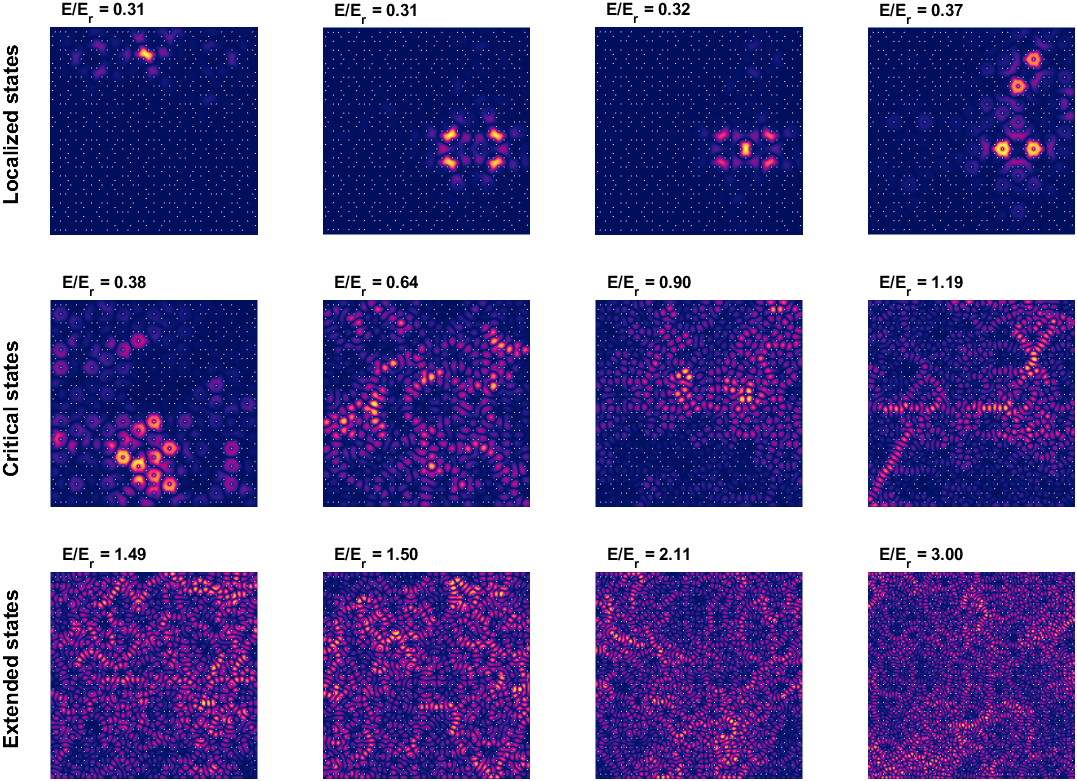}
    \caption{Wavefunction amplitude $|\psi|$ of example localized, critical, and extended states. The energy of each state is indicated. The side length of the system here is $L = 40a$. White transparent dots indicate the locations of the potential maxima.}
    \label{fig.loc_crit_ext_states}
\end{figure*}

The results show that criticality ($D_2<2$) occurs mostly around lower energies and around the spectral pseudogaps. This is in agreement with recent calculations on 8-fold symmetric aperiodic potentials~\cite{Zhu_PRA2024}. To illustrate better the correlation between the critical states and the presence spectral gaps we show in Fig.~\ref{fig.D2_vs_En}c a zoom-in of a region with critical states ($D_2<2$) where clear minigaps appear in the spectrum that could not be discerned in panel (a). At higher energies waves are less-and-less diffracted by the Gaussian potentials at the vertices of the tiling and extended states start becoming dominant with a dimension approaching $D_2 = 2$ (see Fig.~\ref{fig.D2_vs_En}a). To obtain a better visual understanding of how localized, critical, and extended states look like in the Monotile, we plot in Fig.~\ref{fig.loc_crit_ext_states} the wavefunction amplitude $|\psi|$ of example localized states (top row) which we previously filtered out, and critical states (middle row) and extended states (bottom row). 
%Notice that the localized states correspond mostly to $E/E_r\approx 0.3$ which can be seen as a blue dot in the center of the first large pseudogap in Fig.~\ref{fig.D2_vs_En}a. 
The spatial window here is $L = 40a$.

\begin{figure}
    \centering
    \includegraphics[width=\linewidth]{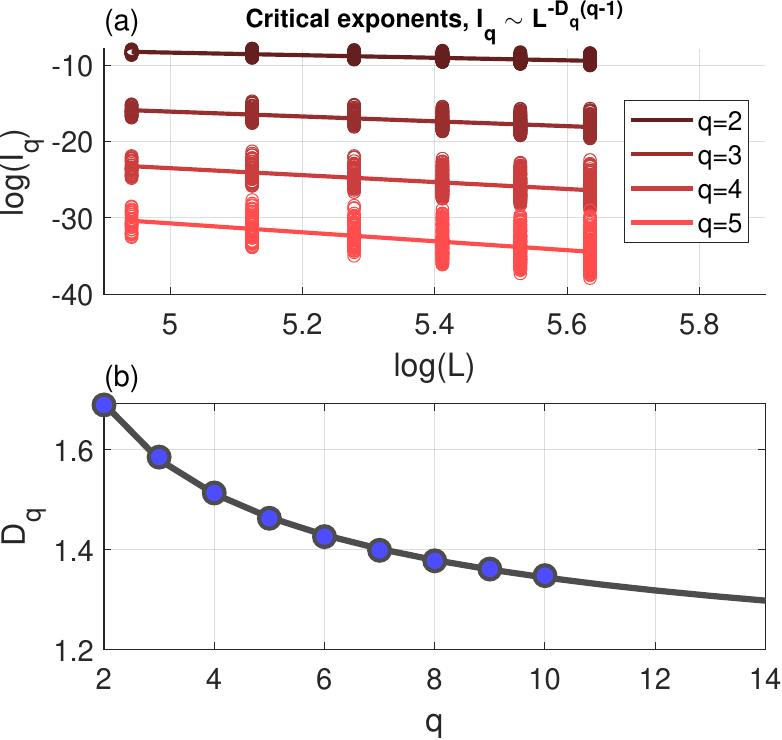}
    \caption{(a) Double logarithmic plot of the moments of the average inverse participation function $\log{(I_q)}$ against system size $\log{(L)}$ for Monotile eigenstates in the energy window $[E-\delta, E+\delta]$ where $E = 1.25 E_r$ and $\delta = 0.2 E_r$. Simple linear regression gives the critical exponent in the scaling law $I_q \sim L^{-D_q(q-1)}$. The obtained generalized dimensions $D_q$ are plotted against the wavefunction probability distribution moments $q$ showing a power law scaling towards $D_\infty = 1$.}
    \label{fig.Dq}
\end{figure}

To further confirm the onset of criticality within the polariton Monotile we show in Fig.~\ref{fig.Dq}a scatter plots of $
\log{(I_q)}$ as a function of system size $\log{(L)}$ for different moments $q$ within an energy window around $E = 1.25 E_r$ and $\delta = 0.2 E_r$ to show more datapoints. A simple linear regression for each moment gives the solid lines whose slope corresponds to the critical exponents $-D_q(q-1)$ from Eq.~\eqref{eq.Dq}. 
%For completeness, we compare the Penrose and the monotile at the same energies and find that the Penrose QC also possesses nontrivial values of the critical exponent indicating criticality. This confirms that the criticality we observe is not exclusive to the monotile and rather a feature of aperiodic systems as expected. 
%Figure~\ref{fig.Dq}b shows the decrease of the slope for higher moments of the wavefunction distribution $\log{(I_q)}$ and 
Figure~\ref{fig.Dq}b plots the obtained slopes up to $q=10$. In good agreement with the theory of critical systems~\cite{Evers_RevModPhys2008} we find that a power law fits the dependence of $D_q$ that approaches an asymptotic value $D_\infty = 1.01$ in agreement with $D_\infty = d/2$ in critical systems~\cite{Lima_2005PRB}.

%Note that if a system possesses a mixture of localized and extended states then $D_q<d$ is not sufficient to determine multifractality.

%Namely, in a 2D system the following limit exists for extended states $\lim_{L\to \infty} I = 1/L^2$ whereas for localized states it becomes $\lim_{L\to \infty} I = \text{const.}$

\section{Transverse localization and transport properties} \label{sec.transport}

Having established the fractal nature of the Monotile, we next investigate how its criticality results in anomalous transport properties~\cite{Yuan_PRB2000, Sbroscia_PRL2020, Zhu_PRA2024, Wang2024} that might be measured in a polariton cavity. Specifically, we explore the spread of an initial polariton wavepacket and identify both superdiffusive and near subdiffusive regimes in the polariton Monotile associated with its fractal spectrum. 

Localization and inhibited transport of spreading linear waves is most famously known to occur in disordered materials, known as the {\em Anderson-localization}~\cite{Evers_RevModPhys2008, Segev2013, Sturges2019} due to scattering from random defects. 
%In two-dimensions, even the slightest presence of disorder will eventually cause waves to become localized. 
Similarly, aperiodic systems and quasicrystals display complex interference between diffracted linear waves that can lead to anomalous transport~\cite{Roche_JMP1997, Levi2011}. This is in sharp contrast to the usual case of ballistic transport, where the width of a freely expanding wavepacket on a regular lattice scales linearly in time $w \sim t$ and the case of random-walk diffusion where it scales as square-root in time $w \sim t^{1/2}$. Anomalous transport refers to scaling $w \sim t^\nu$ where the diffusion exponent $\nu$ can take on other values in the range $0<\nu<1$ with $\nu=0$ corresponding to localization. Anomalous transport occurs in systems that possess a fractal energy spectrum (or singular continuous spectrum) such as for 2D electrons in a magnetic field (Harper model)~\cite{Hiramoto1988}, in disordered materials with a metal-insulator Anderson localization transition, and quasicrystals such as the Fibonacci and Aubry–André model. 
%Similar phenomenon was also observed in incommensurate Moir\'{e} lattices based on two stacked periodic sublattices with a twist angle~\cite{Wang2019}. 
%The contrary are e.g. regular lattices with translational symmetry which only possess extended states and a continuous spectrum that leads to normal ballistic propagation of waves.
%\hs{[Þarf að hugsa aðeins meira um þetta. - HS]}. 
%Interestingly, the interplay between disorder and aperiodic order was found to give rise to counterintuitive effects like disorder-enhanced transport~\cite{Levi2011} where random defects serve as bridges for energy transfer in the fractal spectrum of the QC.

%For this purpose we will focus on the fractal dimension of the second moment $D_2$ \hs{as this is equivalent to the box-counting dimension often used for fractal structures and measures the spread of the wave function over the supporting Hilbert space.} 

The spread of an initial wavepacket $\psi(\mathbf{r},t=0) = \psi_0 = \exp{(-r^2/2w_0^2)}/w_0 \sqrt{\pi}$ in the monotile can be characterized by two measures that capture its anomalous scaling in time~\cite{Yuan_PRB2000}. First, there is the effective spatial width $w_\text{eff}(t)$ mentioned earlier,
\begin{equation} \label{eq.weff}
    w_\text{eff}  =  \left[ \int |\psi(\mathbf{r},t)|^4 \, d\mathbf{r} \right]^{-1/2}   \sim t^{\nu}.
 \end{equation}
%In two dimensions the width of the wavefunction scales as $\nu=1$ for unhindered ballistic transport (i.e collision free propagation) and $\nu=1/2$ for random walk diffusive transport. 
We can define subdiffusive $0<\nu<1/2$ and superdiffusive $1/2<\nu<1$ regimes. When the Hilbert space is multifractal the scaling exponent picks up a nontrivial form $\nu = D_2^\mu/D_2^\Psi$ where $D_2^\mu$ and $D_2^\Psi$ are the average spectral and spatial correlation dimensions of the eigenfunctions making up the initial wavepacket $\psi_0$~\cite{Ketzmerick_PRL1997}. The former governs the temporal scaling of the wavepacket whereas the latter governs the shape of the wavepacket's central region $|\psi| \sim r^{D_2^\psi - 2}$. 
%govern the temporal and spatial power law decay of the wavepacket. 
%In particular, the {\em return probability} of the wavefunction scales as $R(t) = |\langle \psi_0 | \psi(t) \rangle|^2  \sim \lim_{t\to\infty} L^{-D_2^\Psi}$.

The second measure is the temporal autocorrelation function $C(t)$ defined as the smoothed probability of the initial state at time $t$,
\begin{equation} \label{eq.C}
    C(t) = \frac{1}{t} \int_0^{t} R(t') \, dt' \sim t^{-\delta},
 \end{equation}
where $R(t) = |\langle \psi_0 | \psi(t) \rangle|^2$ is the {\em return probability} of the wavefunction. In two dimensions, ballistic transport corresponds to $\delta = 1$ whereas random walk diffusion $C \sim \log{(t)}/t$. The algebraic decay of the autocorrelation function in systems with multifractal spectrum was shown to follow the spectral correlation dimension $\delta = D_2^\mu$ of the local density of states~\cite{Zhong1995, Ketzmerick_PRL1997}. In particular, for a Cantor set spectrum one can relate the spectral exponent to the generalized dimension $D_2^\mu = D_2/d$~\cite{Ketzmerick_PRL1992} and for point spectra systems $\nu,\delta \to 0$ as waves eventually become exponentially localized.

 %In the context of 2D Anderson localization, for any amount of disorder one finds that $\nu \to 0$ as waves eventually become exponentially localized and static $\psi \propto \exp{(-r/\xi_\text{loc})}$ centered at their point of origin. The localization length is defined $\xi_\text{loc} = \ell \exp{(\pi k_\perp \ell/2)}$ with $\ell$ being the mean free path of the particle~\cite{Lee_RevModPhys1985, Evers_RevModPhys2008}. 

%The multifractal structure of the eigenstates manifests as diffusive-like spread of wavepackets developing a power-law tail which we propose can be measured in current experiments.

%The non-Hermitian transport problem~\cite{Eichelkraut2013} including these losses is quite interesting in itself and will be discussed separately below.

%Recent work on optically induced QCs in a photorefractive medium~\cite{Wang2024} has shown that the amount of transverse localization is tied to the degree of discrete rotational symmetry in the QC. That is, high rotational symmetric QCs are more likely to have localized states that can slow down the transverse spreading of wavepackets in the QC. Since the monotile is not rotationally symmetric,
%\hs{[Ég þarf aðeins að hugsa um þetta. Samkvæmt~\cite{Schirmann_PRL2024} er einflísin handhverf en líka með $C_6$ snúningssamhverfu. Hmmm. - HS]}, 
%we expect it to possess more delocalized (extended) modes and show more efficient transport in the plane as compared to e.g. the Penrose QC with its 5-fold rotational symmetry

\begin{figure}
    \centering
    \includegraphics[width=\linewidth]{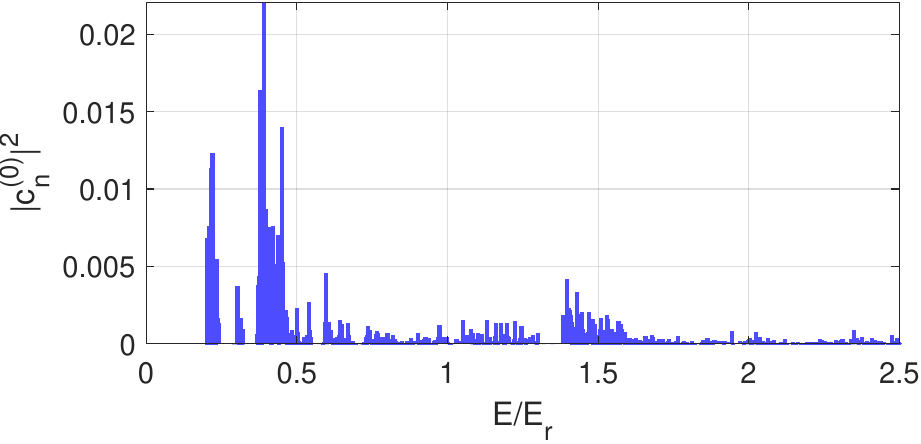}
    \caption{Local density of states of the projected initial wavepacket $\psi_0$ on the eigenstates of the Monotile.}
    \label{fig.proj}
\end{figure}

Rather than calculating $D_2^\mu$ and $D_2^\Psi$ directly from the eigenstates we instead obtain $\nu$ and $\delta$ from the dynamics of the initial wavepacket problem through direct numerical integration of the 2D Schrödinger equation using the split-step fast Fourier transform method. We note that the input Gaussian wavepacket $\psi_0$ can be realized in a polariton system through a resonant injection with a Gaussian optical beam at normal incidence to the cavity. Here, the width of the wavepacket is chosen to larger than the optical lattice pump spots $w_0 = 3 w$ in order to have greater projection on low energy fractal eigenstates instead of extended high energy eigenstates. Figure~\ref{fig.proj} shows the local DOS for this wavepacket with coefficients $c_n^{(0)} = \langle \psi_0 | \psi_n\rangle$ in the spectral decomposition $\psi(t) = \sum_n c_n^{(0)} \psi_n e^{-i E_n t/\hbar}$ for $a=7$ $\mu$m and $V_0 = 1$ meV (same parameters as for Fig.~\ref{fig.D2_vs_En}). This wavepacket achieves a good overlap with critical eigenstates of the Monotile whose average fractality leads to anomalous transport. 
%Notice that the local DOS is highest around the spectral gaps and also shows a rather complex structure which has been found in other 2D fractal systems~\cite{Zhu_PRA2024, Wang2024}.

\begin{figure*}[t]
    \centering
    \includegraphics[width=\linewidth]{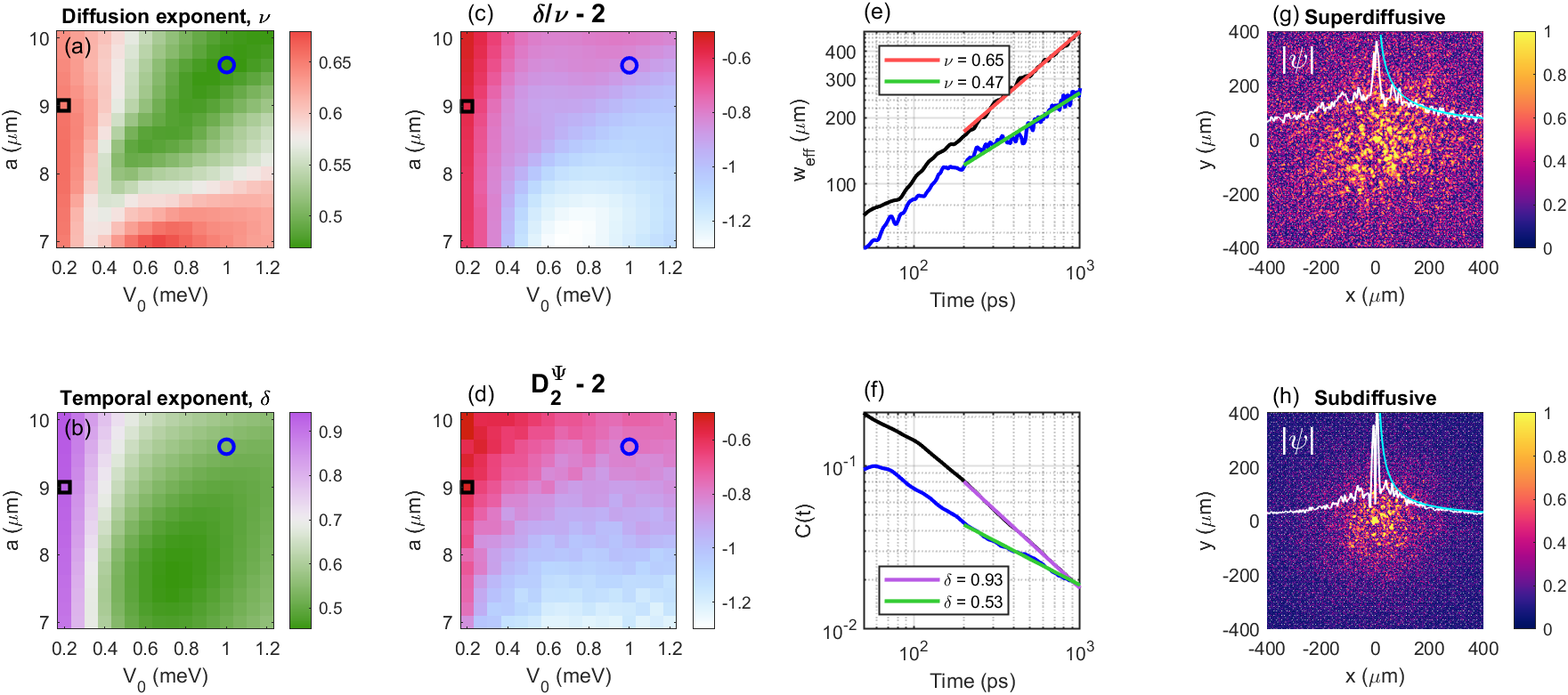}
    \caption{2D Schrödinger simulations of wavepacket spread in the Monotile polariton lattice. (a) Diffusion exponent $\nu$ and (b) temporal correlation exponent $\delta$ obtained after $500$ ps transport in a system of size $L= 100a$. Shape exponent $D_2^\psi$ of the wavepacket's central region at the final time obtained (c) from the temporal and spatial diffusion exponents and (d) from applying a power law fit to $|\psi|$. (e,f) Two example evolution trajectories of $w_\text{eff}$ and $C(t)$ on a log-log plot corresponding to the black square and blue circle in other panels. The slope is obtained through a linear fit at later times which gives $\nu$ and $\delta$, respectively. (g,h) Zoom in of the final time wavefunction amplitude $|\psi|$ for the black square (superdiffusive) and blue circle (subdiffusive) showing a clear difference in spread. A semilog cross-section of the wavefunction is overlaid a solid white curve. The cyan solid curves denote the power law fit $|\psi| \sim r^{D_2^\psi-2}$ to the wavepacket central region.}
    \label{fig.transport1}
\end{figure*}

Figures~\ref{fig.transport1}a and~\ref{fig.transport1}b show the resulting diffusion $\nu$ and temporal correlation $\delta$ exponents over a range of potential amplitudes $V_0$ and ``lattice constant'' $a$. Each pixel corresponds to the same initial condition with $\psi_0$ released at the start. The exponents are obtained by linear fitting of the numerical data on a log-log scale after sufficiently long propagation times. Anomalous transport is evidenced through regimes of both sub-diffusive (deep green) and super-diffusive (red) transport in the diffusion exponent $\nu$.

Assuming that the shape of the wavepacket remains unchanged after initial transients then the following should hold $\delta/\nu = D_2^\psi$~\cite{Ketzmerick_PRL1997}. In order to check this we extract $D_2^\psi-2$ from a power-law fit on the azimuthally averaged envelope $\langle |\psi(\mathbf{r})|\rangle_\varphi$ at the final timestep and compare the results in Figs.~\ref{fig.transport1}c and~\ref{fig.transport1}d showing good agreement. 
%We have chosen the spatial range $r_\text{env}/w_\text{eff} \in [0.2,1]$ to extract $D_2^\psi$. 
%We note that the tails of the wavepacket will assume a stretched exponential with an stretching exponent that is universally tied with the diffusion exponent $\nu$~\cite{Zhong_PRL2001}. 
Deviations between the two methods likely originate from finite propagation time, lack of ensemble averaging, and error from data fitting. Two such example fits in the super- and sub-diffusive regimes (black square and blue circle respectively) are shown in Figs.~\ref{fig.transport1}e and~\ref{fig.transport1}f. The corresponding wavefunctions $|\psi|$ at the final timestep are shown in Figs.~\ref{fig.transport1}g and~\ref{fig.transport1}h. Notice how distorted the wavepacket becomes due to the low symmetry of the Monotile, similar to transport in disordered systems. This is in sharp contrast to wavepackets in rotationally symmetric two-dimensional quasicrystals which still inherit their symmetries~\cite{Freedman2006, Wang2024}
.

\section{Non-Hermitian Monotile transport}
\label{sec:nonherm}
With cavity polaritons being naturally dissipative system requiring external drive we now consider the non-Hermitian transport problem which introduces two new parameters into the Schrödinger equation, 
\begin{equation} \label{eq.schro_NH}
    i \hbar \frac{\partial \psi}{\partial t} = \left[- \frac{\hbar^2 \nabla^2}{2m} + (V_0 + i \kappa)V(x,y) - \frac{i \hbar \gamma}{2} \right] \psi.
\end{equation}
Here, $\gamma$ is the inverse lifetime of the polariton and $\kappa$ quantifies local amplification of polaritons around the potential.

For $\kappa = 0$ and at sufficiently long times it becomes immediately apparent that
\begin{equation}
    C(t) \sim t^{-1}.
\end{equation}
That is, the return probability decays as $R \sim e^{-\gamma t/2}$. However, the effective width $w_\text{eff}$ is not affected by global losses and remains the same as in the Hermitian case. When $\kappa > 0$ things become radically different as the pump spots now effectively mimic non-elastic scatterers. The probability current $\mathbf{j}$ is no longer conserved ($\nabla \cdot \mathbf{j} \neq 0$) and the continuity equation for the eigenstates instead becomes $\nabla \cdot \mathbf{j} =  (2 \kappa V/\hbar - \gamma )|\psi_n|^2$. Wave transport properties are much less studied in non-Hermitian structures that can display some peculiar effects like sudden transition from ballistic to diffusive despite being perfectly ordered~\cite{Eichelkraut2013}.

The effects of making the potentials complex is highly relevant for cavity polaritons as $\kappa$ represents optical gain coming from the pump spots. If $\kappa$ is large enough compared to the losses $\gamma$, then a single complex eigenvalue will cross the lower half of the complex plane to the upper half and $\psi_n \sim e^{+\lambda_n t}$ will diverge at a rate of $\lambda_n$ which is the largest (positive) imaginary part of the complex eigenenergy. This crossing defines the nonequilibrium condensation threshold of polaritons from a mean field perspective~\cite{Deng_RMP2010} (see Section~\ref{sec.cond}).
%In order describe kinetic equilibrium and associated dynamics of the field, the 2D Schrödinger equation must be generalized to a nonlinear form similar to the Maxwell-Bloch equation that describes dynamics between lasing and charge carriers. Here, such a dynamical model for polaritons is referred as the generalized 2D Gross-Pitaevskii equation which we discuss in more detail in the next section. 

For the purpose of this section, we will concern ourselves with the case of $\kappa$ being sufficiently small so that all eigenenergies are in the lower half of the complex plane (below threshold). We will focus on the scaling properties of the wavepacket width as we increase $\kappa$ up to this critical point using a inverse participation ratio defined now as,
\begin{equation}
    \tilde{I}_2  =  \frac{\int |\psi|^{4} d\mathbf{r}}{\left( \int |\psi|^2 \, d\mathbf{r} \right)^2},
\end{equation}
where $w_\text{eff} = \tilde{I}_2^{-1/2} \sim t^\nu$. We set $\hbar \gamma = 0.1$ meV which is typical for planar cavities made of distributed Bragg reflectors.

\begin{figure}
    \centering
    \includegraphics[width=\linewidth]{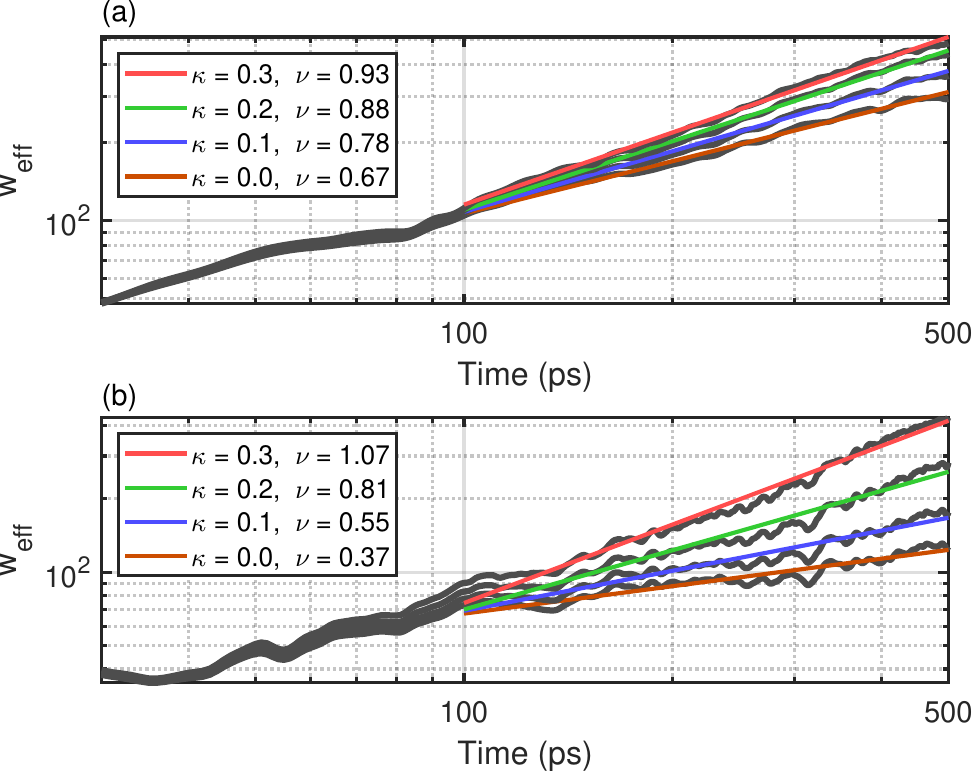}
    \caption{Super- (a) and sub-diffusive (b) scaling of the polariton wavepacket in a non-Hermitian Monotile quasilattice, corresponding to the parameters of the black square and blue circle in Fig.~\ref{fig.transport1}. As the non-Hermitian parameter $\kappa$ increases the diffusion becomes more ballistic as extended waves survive low-energy critical waves.}
    \label{fig.kappa}
\end{figure}

Figure~\ref{fig.kappa} shows a log-log plot of the effective width against time for the same wavepacket $\psi_0$ as in previous section but now following Eq.~\eqref{eq.schro_NH} for varying levels of non-Hermiticity $\kappa$. We choose two example values of potential amplitude $V_0$ and ``lattice spacing'' $a$ that give super- and sub-diffusive behaviors in the hermitian case corresponding to the square and circle markers in Fig.~\ref{fig.transport1}. The results show that as $\kappa$ increases the wavepacket spreads more ballistically with $\nu \to 1$. The reason for this behaviour can be understood intuitively. Low energy waves belonging to critical states have poor overlap with the potential regions (pump spots) and are therefore much less amplified compared to higher energy extended waves. Therefore, the projection of the wavepacket on the critical states decreases faster in time until the dynamics are mostly dictated by the surviving extended states which expand ballistically. We will see in the next section that this is reflected directly in the stable solutions of interacting ballistic polariton condensates in the Monotile.

\section{Mean field dynamics of Monotile polariton fluids} \label{sec.cond}
%In this section we explore the condensation dynamics of polaritons in the critical states  of the monotile.
The single-particle problem discussed in the previous section sets the stage for the many-body problem where we apply the widely successful mean field formalism to describe the coherent fraction of the polariton condensate. As mentioned previously, each condensate is driven above threshold by a tightly focused nonresonant Gaussian pump spot and subsequently forms a macroscopically coherent object---a droplet of light---that emit polaritons laterally in all directions in the cavity plane~\cite{Richard_PRL2005}. The ballistic outflow of coherent polaritons from each condensate gives rise to long range non-Hermitian coupling~\cite{Ohadi_PRX2016, Topfer2020} beyond nearest neighbors~\cite{Alyatkin2020} and spontaneous synchronization into various phase-configurations depending on the pump geometry~\cite{Tosi2012b, Berloff2017, Alyatkin2021}.

Within the mean field formalism where quantum effects are negligible the non-Hermitian Schrödinger equation is generalized to a nonlinear form similar to the Maxwell-Bloch equation that describes dynamics between lasing and charge carriers. The polariton condensate is described as a macroscopic coherent wavefunction $\psi(\mathbf{r},t)$ (i.e. {\it order parameter}) coupled to a background reservoir density of incoherent excitons $n_R(\mathbf{r},t)$ obeying the generalized Gross-Pitaevskii equation~\cite{Deng_RMP2010},
\begin{align} \notag
i\hbar \frac{\partial \psi}{\partial t} &= \bigg[-\frac{\hbar^2 \nabla^2}{2m} + \alpha |\psi|^2 + G\left(n_R + \frac{\eta P}{\Gamma}\right) ... \\ \label{eq.2dgpe}
& +  \frac{i \hbar (Rn_R - \gamma)}{2}\bigg]\psi + f(\mathbf{r},t) \\ \notag
    \partial_t n_R &= -(\Gamma + R|\psi|^2)n_R + P(\mathbf{r}).
\end{align}
Here, $\alpha$ is the interaction strength between polaritons in the condensate, $G$ is the interaction strength between polaritons and reservoir excitons, $\Gamma$ is the reservoir exciton decay rate, $\gamma$ is the decay rate of the lower polaritons, $R$ is the rate of stimulated scattering of polaritons into the condensate from the exciton reservoir, and $\eta$ is a phenomenological constant accounting for additional blueshift due to charge carriers and high-momentum exciton background. The nonresonant pump profile $P(\mathbf{r})$ is constructed through a superposition of Gaussian beams just like in Eq.~\eqref{eq.V} but with $V_0 \to P_0$ where $P_0$ is the pump power density. For small condensate densities one can write the steady state solution of the excitons as $n_R \approx P/\Gamma$ which means that the optical potential amplitude in Eq.~\eqref{eq.V} can be related to the 2DGPE parameters through $V_0 = G P_0(1+\eta)/\Gamma$. The gain parameter can be similarly written $\kappa = \hbar R P_0/2\Gamma$. We additionally include a resonant driving term $f(\mathbf{r},t)$ which corresponds to an additional coherent laser drive at normal incidence that can be tuned in resonance with the polariton modes. 

\subsection{Nonresonantly driven condensation}
Ballistic polariton condensates are by definition extended wave-objects where coherent polariton waves radiate in all directions away from the pump spots. It is therefore only to be expected that when the system is driven above threshold $P_0>P_\text{thr}$ the resulting dynamics will be dominated by extended waves as was recently reported for polaritons in an optical Penrose tiling~\cite{Alyatkin2025} and Monotile~\cite{alyatkin2026}. In fact, ballistic polariton condensation usually takes place in modes characterized by wavevectors determined by their local pumping potential $k_c \approx \sqrt{2 m V_0}/\hbar$~\cite{Wouters_PRB2008}. According to our earlier results (see Fig.~\ref{fig.D2_vs_En}) critical states appear at low energies in the Monotile in the optical polariton potential. Even if the condensate could be encouraged to populate low energy states in the same range as critical states (e.g. by pumping hard enough to facilitate energy relaxation of polaritons) there is no guarantee it will pick a critical state to condense in. In fact, the strong cooperativity and high-field seeking behaviour of polariton condensates~\cite{Berloff2017, Alyatkin2021}, similar to coupled lasers, will most likely result in a condensate of maximum extend in order to optimize its gain.

\begin{figure}
    \centering
    \includegraphics[width=0.9\linewidth]{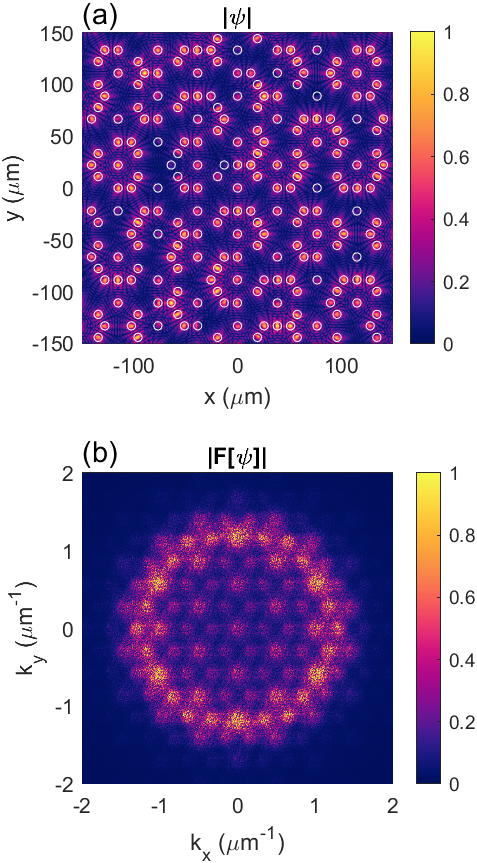}
    \caption{Zoom in of the steady state solution of the polariton Monotile driven above condensation threshold in (a) real space and (b) momentum space. White circles denote the locations of the pumps. The ballistic nature of the polariton condensates pumped by tightly focused Gaussian spots is captured in the vivid interference fringes between the condensates. Parameters are similar to those used in Section~\ref{sec.transport}.}
    \label{fig.2dgpe}
\end{figure}

To illustrate this point, we show in Fig.~\ref{fig.2dgpe} a zoom-in of typical ballistic condensate network above threshold in  real- and momentum-space obtained through direct intergration of Eq.~\eqref{eq.2dgpe} with random white-noise initial conditions. We have selected parameters that are same as those used in previous experiments~\cite{Alyatkin2025}. Here, the closest distance between the pump spots is chosen to be $a = 12.8$ $\mu$m to show more clearly the interference fringes between the condensates. The white circle markers denote the location of the pumps. The rich interference fringes between the condensates is a hallmark of their ballistic nature reflecting their extended macroscopic coherence. Notice that for some pump markers in Fig.~\ref{fig.2dgpe}a the condensate has a very weak amplitude due to complex interference between neighbours.

Performing scaling analysis [Eq.~\eqref{eq.Dq}] on the obtained condensate solution for varying system size we obtain $D_2 \approx 2$ in agreement with the condensate belonging to extended modes. The momentum space (Fourier transformed) signal in Fig.~\ref{fig.2dgpe}b shows the $C_6$ rotational symmetry structure inherited from the underlying tetrille lattice~\cite{Socolar_PRB2023}. It is beyond the scope of the current study to analyze the complex synchronization patterns in the ballistic polariton condensate Monotile which were recently reported~\cite{alyatkin2026}. This will be left for a future study and we focus instead on transport properties.
%Currently, we are interested in how the critical states of the Monotile manifest in the nonlinear dynamics of polariton quantum fluids. 
%Instead we consider resonant injection of polaritons in Section~\ref{sec.res} which forms a more controlled method of generating polariton quantum fluids at the sacrifice of spontaneous symmetry breaking. 
%

\subsection{Nonlinear anomalous transport in a pulsed polariton quantum fluid} \label{sec.res}
Here, we propose that the anomalous transport properties of the polariton Monotile can be probed using a resonant picosecond pulse which excites the critical states. This scenario is similar to the starting wavepacket transport calculations discussed in section~\ref{sec.transport} but we now solve the full dynamics of the 2DGPE. This forms a more controlled method of generating polariton quantum fluids with specific parameters at the sacrifice of spontaneous symmetry breaking.

The resonant driving term is taken to be a Gaussian picosecond pulsed beam with spatial width $w_0 = 3w$ (same as in our initial wavepacket analysis) and temporal width $w_t = 2$ ps, and amplitude $f_0$, 
\begin{equation}
%f(\mathbf{r},t) = \exp{(-r^2/2w_0^2)} \cdot \exp{(-(t-t_0)^2/2w_t^2)}
f(\mathbf{r},t) = f_0 e^{-r^2/2w_0^2} \cdot e^{-(t-t_0)^2/2w_t^2}.
\end{equation}
The temporal width is chosen to be short enough to excite mostly the critical states within energies $\hbar/(2 w_t E_r) <  1.5$. Here, $t_0 = 3w_t$ is chosen so the that the entire pulse starts is within the simulation time window. 

\begin{figure}[t]
    \centering
    \includegraphics[width=\linewidth]{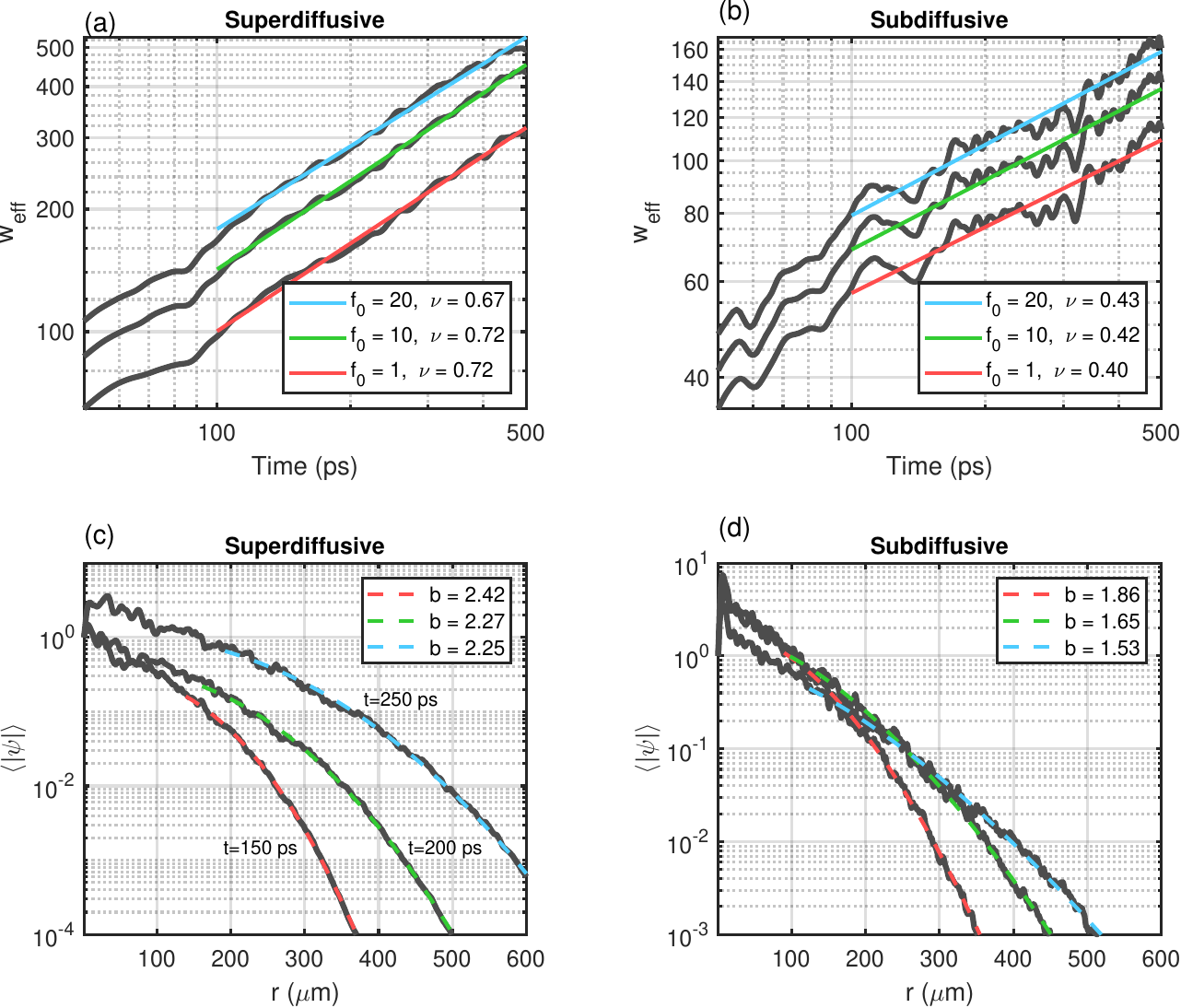}
    \caption{Anomalous transport of the pulsed polariton fluid. Effective width $w_\text{eff}$ is shown for different drive amplitudes in the (a) superdiffusive and (b) subdiffusive regime. Defocusing nonlinear effects at early times increase the width of the polariton fluid but do not play a strong role at long times where we recover the same superdiffusive scaling exponent $\nu \approx 0.7$ and $\nu \approx 0.4$. (c,d) Semilog plot of the scaled envelope at different times for a low power pulse ($f_0=1$) with stretched exponential fits $|\psi| \sim e^{-(r/a)^b}$ (colored dashed curves).}
    \label{fig.2dgpe_wp}
\end{figure}

The increase of polariton fluid width $w_\text{eff}$ is shown in Fig.~\ref{fig.2dgpe_wp} for different drive amplitudes $f_0$ in the superdiffusive and subdiffusive case. The defocusing nonlinearity $\alpha |\psi|^2>0$ in Eq.~\eqref{eq.2dgpe} causes an abrupt increase in the width of the polariton fluid due to its repulsive action leading to a vertical shift in the trajectories of $w_\text{eff}$. However, due to the constant loss rate $\gamma$, the polariton fluid quickly dissipates at longer times where we recover a similar diffusion exponent across all ranges of power. The retrieved values of $\nu$ correspond well to our results on the wavepacket spread in the single-particle picture in Fig.~\ref{fig.transport1} confirming anomalous transport. 

Another method of verifying anomalous transport in the pulsed Monotile comes from looking at the wavefront of the expanding polariton distribution, whose shape is tied to the dissipation exponent $\nu$. Figures~\ref{fig.2dgpe_wp}c and~\ref{fig.2dgpe_wp}d show a semilog plot of the scaled polariton fluid envelope $\langle |\psi(r,t)| \rangle_\varphi/|\psi(0,t)|$ at different times for a low power pulse ($f_0=1$). We fit a stretched exponential $|\psi| \sim e^{-(r/a)^b}$ to the numerical data in the range $r_\text{fit}/w_\text{eff} \in [1,5]$ revealing superdiffusive transport with $b>2$ and subdiffusive transport $b<2$ in agreement with established theory on anomalous diffusion~\cite{Zhong_PRL2001}. We find that the established relation between the shape and diffusion exponent $b = 1/(1-\nu)$~\cite{Zhong_PRL2001} holds nicely for the subdiffusive case $1/(1-0.4) = 1.67$ but a stronger mismatch appears in the superdiffusive case $1/(1-0.72) = 3.57$ which comes from the finite spatial and temporal width of the pulse and our simulation grid which we have checked on less computationally intensive Fibonacci chains. Nevertheless, clear differences between the two transport regimes in the Monotile can be achieved in a pulsed experiment on polaritons by simply tuning two parameters of the optical potential, $a$ and $V_0$.  %This deviation comes most likely from the finite-width pulsed injection instead of an ideal delta-function initial condition.

\section{Conclusions}
We have performed extensive numerical calculations and analysis on wave transport properties of ``the Hat'' monotile quasilattice, the latest member into the family of aperiodic tilings~\cite{Smith2024}. Our system is motivated by recent experimental demonstrations of spontaneous long-range coherence in an optical Penrose~\cite{Alyatkin2025} and Monotile~\cite{alyatkin2026} quasilattices of polariton condensates. Nearly all studies on exciton-polaritons in aperiodic systems have focused on 1D systems with little exploration into 2D system which we have aimed to fill. 

Our main results underpin the quantum criticality and fractal dimensionality of the polariton Monotile system. The main observables associated with these features is reflected in the anomalous transport of polaritons which can be readily measured through standard optical cavity setups. We identify regimes of superdiffusive and near subdiffusive wavepacket spread connected with the Monotile generalized fractal dimension. We extend our anomalous wave-transport results to the non-Hermitian and mean-field equations of motion which describe the dynamics of a non-resonantly and resonantly driven macroscopic polariton quantum fluid (i.e. condensate). Our findings open perspectives on studying critical phenomena and anomalous transport in all-optically versatile polariton systems where many parameters of the potential can be adjusted~\cite{Alyatkin2025} and even modulated in time~\cite{delValleInclanRedondo2024}. 

While optical polariton networks are generally limited to around 100 pump spots above condensation threshold~\cite{Alyatkin2021} it would be possible to scale up to 1000 spots using a pump-probe setup where a low power pump pulse creates the potential below threshold and a probe sends in a subsequent polariton wavepacket. It is worth noting that our results are not exclusive to the Monotile and we have also discovered similar results in the Penrose tiling. It still remains a question what are the unique advantages of the Monotile against other aperiodic systems which we aim to uncover in future work.

The next step of our investigation will involve stochastic discrete nonlinear networks (vertex models) of coupled polariton condensates in a Monotile graph geometry with distance-dependent couplings. Such condensate models can provide insight into the buildup of correlations and fractal structure in the elementary (Bogoliubov) excitations. In particular, it would be interesting to explore the energy structure of Monotile waves in a distance-dependent tight-binding model since previous efforts only considered same strength coupling between all near neighbours~\cite{Schirmann_PRL2024}. This would shed further light on the role of the Monotile as a topological insulator~\cite{Schirmann_PRL2024} and for generation of anomalous mobility edges~\cite{Duncan_PRB2024}. Lastly, a chiral aperiodic monotile~\cite{Smith2024_b} might display interesting polarization properties when considering the polariton spin in conjunction with photonic spin-orbit effects and polarized pumping of definite handedness.

\section*{Acknowledgements}
V.K.D acknowledges the Icelandic Research Fund (Rann\'{i}s), grant No. 239552-051. H.S. acknowledges the National Science Center, Poland, project No. 2024/55/B/ST3/02954.



\begin{thebibliography}{57}%
\makeatletter
\providecommand \@ifxundefined [1]{%
 \@ifx{#1\undefined}
}%
\providecommand \@ifnum [1]{%
 \ifnum #1\expandafter \@firstoftwo
 \else \expandafter \@secondoftwo
 \fi
}%
\providecommand \@ifx [1]{%
 \ifx #1\expandafter \@firstoftwo
 \else \expandafter \@secondoftwo
 \fi
}%
\providecommand \natexlab [1]{#1}%
\providecommand \enquote  [1]{``#1''}%
\providecommand \bibnamefont  [1]{#1}%
\providecommand \bibfnamefont [1]{#1}%
\providecommand \citenamefont [1]{#1}%
\providecommand \href@noop [0]{\@secondoftwo}%
\providecommand \href [0]{\begingroup \@sanitize@url \@href}%
\providecommand \@href[1]{\@@startlink{#1}\@@href}%
\providecommand \@@href[1]{\endgroup#1\@@endlink}%
\providecommand \@sanitize@url [0]{\catcode `\\12\catcode `\$12\catcode
  `\&12\catcode `\#12\catcode `\^12\catcode `\_12\catcode `\%12\relax}%
\providecommand \@@startlink[1]{}%
\providecommand \@@endlink[0]{}%
\providecommand \url  [0]{\begingroup\@sanitize@url \@url }%
\providecommand \@url [1]{\endgroup\@href {#1}{\urlprefix }}%
\providecommand \urlprefix  [0]{URL }%
\providecommand \Eprint [0]{\href }%
\providecommand \doibase [0]{https://doi.org/}%
\providecommand \selectlanguage [0]{\@gobble}%
\providecommand \bibinfo  [0]{\@secondoftwo}%
\providecommand \bibfield  [0]{\@secondoftwo}%
\providecommand \translation [1]{[#1]}%
\providecommand \BibitemOpen [0]{}%
\providecommand \bibitemStop [0]{}%
\providecommand \bibitemNoStop [0]{.\EOS\space}%
\providecommand \EOS [0]{\spacefactor3000\relax}%
\providecommand \BibitemShut  [1]{\csname bibitem#1\endcsname}%
\let\auto@bib@innerbib\@empty
%</preamble>
\bibitem [{\citenamefont {Jens-Boie~Suck}(2002)}]{Suck2002}%
  \BibitemOpen
  \bibfield  {author} {\bibinfo {author} {\bibfnamefont {P.~H.}\ \bibnamefont
  {Jens-Boie~Suck}, \bibfnamefont {Michael~Schreiber}},\ }\href
  {https://doi.org/10.1007/978-3-662-05028-6} {\emph {\bibinfo {title}
  {Quasicrystals: An Introduction to Structure, Physical Properties and
  Applications}}}\ (\bibinfo  {publisher} {Springer Berlin Heidelberg},\
  \bibinfo {year} {2002})\BibitemShut {NoStop}%
\bibitem [{\citenamefont {Steurer}(2004)}]{Steurer2004}%
  \BibitemOpen
  \bibfield  {author} {\bibinfo {author} {\bibfnamefont {W.}~\bibnamefont
  {Steurer}},\ }\bibfield  {title} {\bibinfo {title} {Twenty years of structure
  research on quasicrystals. part i. pentagonal, octagonal, decagonal and
  dodecagonal quasicrystals},\ }\href
  {https://doi.org/10.1524/zkri.219.7.391.35643} {\bibfield  {journal}
  {\bibinfo  {journal} {Zeitschrift f\"{u}r Kristallographie - Crystalline
  Materials}\ }\textbf {\bibinfo {volume} {219}},\ \bibinfo {pages} {391–446}
  (\bibinfo {year} {2004})}\BibitemShut {NoStop}%
\bibitem [{\citenamefont {Hiramoto}\ and\ \citenamefont
  {Abe}(1988)}]{Hiramoto1988}%
  \BibitemOpen
  \bibfield  {author} {\bibinfo {author} {\bibfnamefont {H.}~\bibnamefont
  {Hiramoto}}\ and\ \bibinfo {author} {\bibfnamefont {S.}~\bibnamefont {Abe}},\
  }\bibfield  {title} {\bibinfo {title} {Dynamics of an electron in
  quasiperiodic systems. ii. harper’s model},\ }\href
  {https://doi.org/10.1143/jpsj.57.1365} {\bibfield  {journal} {\bibinfo
  {journal} {Journal of the Physical Society of Japan}\ }\textbf {\bibinfo
  {volume} {57}},\ \bibinfo {pages} {1365–1371} (\bibinfo {year}
  {1988})}\BibitemShut {NoStop}%
\bibitem [{\citenamefont {Roche}\ \emph {et~al.}(1997)\citenamefont {Roche},
  \citenamefont {Trambly~de Laissardière},\ and\ \citenamefont
  {Mayou}}]{Roche_JMP1997}%
  \BibitemOpen
  \bibfield  {author} {\bibinfo {author} {\bibfnamefont {S.}~\bibnamefont
  {Roche}}, \bibinfo {author} {\bibfnamefont {G.}~\bibnamefont {Trambly~de
  Laissardière}},\ and\ \bibinfo {author} {\bibfnamefont {D.}~\bibnamefont
  {Mayou}},\ }\bibfield  {title} {\bibinfo {title} {Electronic transport
  properties of quasicrystals},\ }\href {https://doi.org/10.1063/1.531914}
  {\bibfield  {journal} {\bibinfo  {journal} {Journal of Mathematical Physics}\
  }\textbf {\bibinfo {volume} {38}},\ \bibinfo {pages} {1794} (\bibinfo {year}
  {1997})}\BibitemShut {NoStop}%
\bibitem [{\citenamefont {Vardeny}\ \emph {et~al.}(2013)\citenamefont
  {Vardeny}, \citenamefont {Nahata},\ and\ \citenamefont
  {Agrawal}}]{Vardeny2013}%
  \BibitemOpen
  \bibfield  {author} {\bibinfo {author} {\bibfnamefont {Z.~V.}\ \bibnamefont
  {Vardeny}}, \bibinfo {author} {\bibfnamefont {A.}~\bibnamefont {Nahata}},\
  and\ \bibinfo {author} {\bibfnamefont {A.}~\bibnamefont {Agrawal}},\
  }\bibfield  {title} {\bibinfo {title} {Optics of photonic quasicrystals},\
  }\href {https://doi.org/10.1038/nphoton.2012.343} {\bibfield  {journal}
  {\bibinfo  {journal} {Nature Photonics}\ }\textbf {\bibinfo {volume} {7}},\
  \bibinfo {pages} {177–187} (\bibinfo {year} {2013})}\BibitemShut {NoStop}%
\bibitem [{\citenamefont {Viebahn}\ \emph {et~al.}(2019)\citenamefont
  {Viebahn}, \citenamefont {Sbroscia}, \citenamefont {Carter}, \citenamefont
  {Yu},\ and\ \citenamefont {Schneider}}]{Viebahn_PRL2019}%
  \BibitemOpen
  \bibfield  {author} {\bibinfo {author} {\bibfnamefont {K.}~\bibnamefont
  {Viebahn}}, \bibinfo {author} {\bibfnamefont {M.}~\bibnamefont {Sbroscia}},
  \bibinfo {author} {\bibfnamefont {E.}~\bibnamefont {Carter}}, \bibinfo
  {author} {\bibfnamefont {J.-C.}\ \bibnamefont {Yu}},\ and\ \bibinfo {author}
  {\bibfnamefont {U.}~\bibnamefont {Schneider}},\ }\bibfield  {title} {\bibinfo
  {title} {Matter-wave diffraction from a quasicrystalline optical lattice},\
  }\href {https://doi.org/10.1103/PhysRevLett.122.110404} {\bibfield  {journal}
  {\bibinfo  {journal} {Phys. Rev. Lett.}\ }\textbf {\bibinfo {volume} {122}},\
  \bibinfo {pages} {110404} (\bibinfo {year} {2019})}\BibitemShut {NoStop}%
\bibitem [{\citenamefont {Sbroscia}\ \emph {et~al.}(2020)\citenamefont
  {Sbroscia}, \citenamefont {Viebahn}, \citenamefont {Carter}, \citenamefont
  {Yu}, \citenamefont {Gaunt},\ and\ \citenamefont
  {Schneider}}]{Sbroscia_PRL2020}%
  \BibitemOpen
  \bibfield  {author} {\bibinfo {author} {\bibfnamefont {M.}~\bibnamefont
  {Sbroscia}}, \bibinfo {author} {\bibfnamefont {K.}~\bibnamefont {Viebahn}},
  \bibinfo {author} {\bibfnamefont {E.}~\bibnamefont {Carter}}, \bibinfo
  {author} {\bibfnamefont {J.-C.}\ \bibnamefont {Yu}}, \bibinfo {author}
  {\bibfnamefont {A.}~\bibnamefont {Gaunt}},\ and\ \bibinfo {author}
  {\bibfnamefont {U.}~\bibnamefont {Schneider}},\ }\bibfield  {title} {\bibinfo
  {title} {Observing localization in a 2d quasicrystalline optical lattice},\
  }\href {https://doi.org/10.1103/PhysRevLett.125.200604} {\bibfield  {journal}
  {\bibinfo  {journal} {Phys. Rev. Lett.}\ }\textbf {\bibinfo {volume} {125}},\
  \bibinfo {pages} {200604} (\bibinfo {year} {2020})}\BibitemShut {NoStop}%
\bibitem [{\citenamefont {Yu}\ \emph {et~al.}(2024)\citenamefont {Yu},
  \citenamefont {Bhave}, \citenamefont {Reeve}, \citenamefont {Song},\ and\
  \citenamefont {Schneider}}]{Yu2024}%
  \BibitemOpen
  \bibfield  {author} {\bibinfo {author} {\bibfnamefont {J.-C.}\ \bibnamefont
  {Yu}}, \bibinfo {author} {\bibfnamefont {S.}~\bibnamefont {Bhave}}, \bibinfo
  {author} {\bibfnamefont {L.}~\bibnamefont {Reeve}}, \bibinfo {author}
  {\bibfnamefont {B.}~\bibnamefont {Song}},\ and\ \bibinfo {author}
  {\bibfnamefont {U.}~\bibnamefont {Schneider}},\ }\bibfield  {title} {\bibinfo
  {title} {Observing the two-dimensional bose glass in an optical
  quasicrystal},\ }\href {https://doi.org/10.1038/s41586-024-07875-2}
  {\bibfield  {journal} {\bibinfo  {journal} {Nature}\ }\textbf {\bibinfo
  {volume} {633}},\ \bibinfo {pages} {338–343} (\bibinfo {year}
  {2024})}\BibitemShut {NoStop}%
\bibitem [{\citenamefont {Watanabe}\ \emph {et~al.}(2021)\citenamefont
  {Watanabe}, \citenamefont {Bhat}, \citenamefont {Baumgaertl}, \citenamefont
  {Hamdi},\ and\ \citenamefont {Grundler}}]{Watanabe2021}%
  \BibitemOpen
  \bibfield  {author} {\bibinfo {author} {\bibfnamefont {S.}~\bibnamefont
  {Watanabe}}, \bibinfo {author} {\bibfnamefont {V.~S.}\ \bibnamefont {Bhat}},
  \bibinfo {author} {\bibfnamefont {K.}~\bibnamefont {Baumgaertl}}, \bibinfo
  {author} {\bibfnamefont {M.}~\bibnamefont {Hamdi}},\ and\ \bibinfo {author}
  {\bibfnamefont {D.}~\bibnamefont {Grundler}},\ }\bibfield  {title} {\bibinfo
  {title} {Direct observation of multiband transport in magnonic penrose
  quasicrystals via broadband and phase-resolved spectroscopy},\ }\href
  {https://doi.org/10.1126/sciadv.abg3771} {\bibfield  {journal} {\bibinfo
  {journal} {Science Advances}\ }\textbf {\bibinfo {volume} {7}},\ \bibinfo
  {pages} {eabg3771} (\bibinfo {year} {2021})}\BibitemShut {NoStop}%
\bibitem [{\citenamefont {Verre}\ \emph {et~al.}(2014)\citenamefont {Verre},
  \citenamefont {Antosiewicz}, \citenamefont {Svedendahl}, \citenamefont
  {Lodewijks}, \citenamefont {Shegai},\ and\ \citenamefont
  {K\"{a}ll}}]{Verre2014}%
  \BibitemOpen
  \bibfield  {author} {\bibinfo {author} {\bibfnamefont {R.}~\bibnamefont
  {Verre}}, \bibinfo {author} {\bibfnamefont {T.~J.}\ \bibnamefont
  {Antosiewicz}}, \bibinfo {author} {\bibfnamefont {M.}~\bibnamefont
  {Svedendahl}}, \bibinfo {author} {\bibfnamefont {K.}~\bibnamefont
  {Lodewijks}}, \bibinfo {author} {\bibfnamefont {T.}~\bibnamefont {Shegai}},\
  and\ \bibinfo {author} {\bibfnamefont {M.}~\bibnamefont {K\"{a}ll}},\
  }\bibfield  {title} {\bibinfo {title} {Quasi-isotropic surface plasmon
  polariton generation through near-field coupling to a penrose pattern of
  silver nanoparticles},\ }\href {https://doi.org/10.1021/nn503195n} {\bibfield
   {journal} {\bibinfo  {journal} {ACS Nano}\ }\textbf {\bibinfo {volume}
  {8}},\ \bibinfo {pages} {9286–9294} (\bibinfo {year} {2014})}\BibitemShut
  {NoStop}%
\bibitem [{\citenamefont {Tsesses}\ \emph {et~al.}(2025)\citenamefont
  {Tsesses}, \citenamefont {Dreher}, \citenamefont {Janoschka}, \citenamefont
  {Neuhaus}, \citenamefont {Cohen}, \citenamefont {Meiler}, \citenamefont
  {Bucher}, \citenamefont {Sapir}, \citenamefont {Frank}, \citenamefont
  {Davis}, \citenamefont {Meyer~zu Heringdorf}, \citenamefont {Giessen},\ and\
  \citenamefont {Bartal}}]{Tsesses2025}%
  \BibitemOpen
  \bibfield  {author} {\bibinfo {author} {\bibfnamefont {S.}~\bibnamefont
  {Tsesses}}, \bibinfo {author} {\bibfnamefont {P.}~\bibnamefont {Dreher}},
  \bibinfo {author} {\bibfnamefont {D.}~\bibnamefont {Janoschka}}, \bibinfo
  {author} {\bibfnamefont {A.}~\bibnamefont {Neuhaus}}, \bibinfo {author}
  {\bibfnamefont {K.}~\bibnamefont {Cohen}}, \bibinfo {author} {\bibfnamefont
  {T.~C.}\ \bibnamefont {Meiler}}, \bibinfo {author} {\bibfnamefont
  {T.}~\bibnamefont {Bucher}}, \bibinfo {author} {\bibfnamefont
  {S.}~\bibnamefont {Sapir}}, \bibinfo {author} {\bibfnamefont
  {B.}~\bibnamefont {Frank}}, \bibinfo {author} {\bibfnamefont {T.~J.}\
  \bibnamefont {Davis}}, \bibinfo {author} {\bibfnamefont {F.}~\bibnamefont
  {Meyer~zu Heringdorf}}, \bibinfo {author} {\bibfnamefont {H.}~\bibnamefont
  {Giessen}},\ and\ \bibinfo {author} {\bibfnamefont {G.}~\bibnamefont
  {Bartal}},\ }\bibfield  {title} {\bibinfo {title} {Four-dimensional conserved
  topological charge vectors in plasmonic quasicrystals},\ }\href
  {https://doi.org/10.1126/science.adt2495} {\bibfield  {journal} {\bibinfo
  {journal} {Science}\ }\textbf {\bibinfo {volume} {387}},\ \bibinfo {pages}
  {644–648} (\bibinfo {year} {2025})}\BibitemShut {NoStop}%
\bibitem [{\citenamefont {Freedman}\ \emph {et~al.}(2006)\citenamefont
  {Freedman}, \citenamefont {Bartal}, \citenamefont {Segev}, \citenamefont
  {Lifshitz}, \citenamefont {Christodoulides},\ and\ \citenamefont
  {Fleischer}}]{Freedman2006}%
  \BibitemOpen
  \bibfield  {author} {\bibinfo {author} {\bibfnamefont {B.}~\bibnamefont
  {Freedman}}, \bibinfo {author} {\bibfnamefont {G.}~\bibnamefont {Bartal}},
  \bibinfo {author} {\bibfnamefont {M.}~\bibnamefont {Segev}}, \bibinfo
  {author} {\bibfnamefont {R.}~\bibnamefont {Lifshitz}}, \bibinfo {author}
  {\bibfnamefont {D.~N.}\ \bibnamefont {Christodoulides}},\ and\ \bibinfo
  {author} {\bibfnamefont {J.~W.}\ \bibnamefont {Fleischer}},\ }\bibfield
  {title} {\bibinfo {title} {Wave and defect dynamics in nonlinear photonic
  quasicrystals},\ }\href {https://doi.org/10.1038/nature04722} {\bibfield
  {journal} {\bibinfo  {journal} {Nature}\ }\textbf {\bibinfo {volume} {440}},\
  \bibinfo {pages} {1166–1169} (\bibinfo {year} {2006})}\BibitemShut
  {NoStop}%
\bibitem [{\citenamefont {Levi}\ \emph {et~al.}(2011)\citenamefont {Levi},
  \citenamefont {Rechtsman}, \citenamefont {Freedman}, \citenamefont
  {Schwartz}, \citenamefont {Manela},\ and\ \citenamefont {Segev}}]{Levi2011}%
  \BibitemOpen
  \bibfield  {author} {\bibinfo {author} {\bibfnamefont {L.}~\bibnamefont
  {Levi}}, \bibinfo {author} {\bibfnamefont {M.}~\bibnamefont {Rechtsman}},
  \bibinfo {author} {\bibfnamefont {B.}~\bibnamefont {Freedman}}, \bibinfo
  {author} {\bibfnamefont {T.}~\bibnamefont {Schwartz}}, \bibinfo {author}
  {\bibfnamefont {O.}~\bibnamefont {Manela}},\ and\ \bibinfo {author}
  {\bibfnamefont {M.}~\bibnamefont {Segev}},\ }\bibfield  {title} {\bibinfo
  {title} {Disorder-enhanced transport in photonic quasicrystals},\ }\href
  {https://doi.org/10.1126/science.1202977} {\bibfield  {journal} {\bibinfo
  {journal} {Science}\ }\textbf {\bibinfo {volume} {332}},\ \bibinfo {pages}
  {1541–1544} (\bibinfo {year} {2011})}\BibitemShut {NoStop}%
\bibitem [{\citenamefont {Xu}\ \emph {et~al.}(2021)\citenamefont {Xu},
  \citenamefont {Wang}, \citenamefont {Chen}, \citenamefont {Smith},\ and\
  \citenamefont {Jin}}]{Xu2021}%
  \BibitemOpen
  \bibfield  {author} {\bibinfo {author} {\bibfnamefont {X.-Y.}\ \bibnamefont
  {Xu}}, \bibinfo {author} {\bibfnamefont {X.-W.}\ \bibnamefont {Wang}},
  \bibinfo {author} {\bibfnamefont {D.-Y.}\ \bibnamefont {Chen}}, \bibinfo
  {author} {\bibfnamefont {C.~M.}\ \bibnamefont {Smith}},\ and\ \bibinfo
  {author} {\bibfnamefont {X.-M.}\ \bibnamefont {Jin}},\ }\bibfield  {title}
  {\bibinfo {title} {Quantum transport in fractal networks},\ }\href
  {https://doi.org/10.1038/s41566-021-00845-4} {\bibfield  {journal} {\bibinfo
  {journal} {Nature Photonics}\ }\textbf {\bibinfo {volume} {15}},\ \bibinfo
  {pages} {703–710} (\bibinfo {year} {2021})}\BibitemShut {NoStop}%
\bibitem [{\citenamefont {Wang}\ \emph {et~al.}(2024)\citenamefont {Wang},
  \citenamefont {Fu}, \citenamefont {Konotop}, \citenamefont {Kartashov},\ and\
  \citenamefont {Ye}}]{Wang2024}%
  \BibitemOpen
  \bibfield  {author} {\bibinfo {author} {\bibfnamefont {P.}~\bibnamefont
  {Wang}}, \bibinfo {author} {\bibfnamefont {Q.}~\bibnamefont {Fu}}, \bibinfo
  {author} {\bibfnamefont {V.~V.}\ \bibnamefont {Konotop}}, \bibinfo {author}
  {\bibfnamefont {Y.~V.}\ \bibnamefont {Kartashov}},\ and\ \bibinfo {author}
  {\bibfnamefont {F.}~\bibnamefont {Ye}},\ }\bibfield  {title} {\bibinfo
  {title} {Observation of localization of light in linear photonic
  quasicrystals with diverse rotational symmetries},\ }\href
  {https://doi.org/10.1038/s41566-023-01350-6} {\bibfield  {journal} {\bibinfo
  {journal} {Nature Photonics}\ }\textbf {\bibinfo {volume} {18}},\ \bibinfo
  {pages} {224–229} (\bibinfo {year} {2024})}\BibitemShut {NoStop}%
\bibitem [{\citenamefont {Shi}\ \emph {et~al.}(2024)\citenamefont {Shi},
  \citenamefont {Peng}, \citenamefont {Jiang}, \citenamefont {Peng},
  \citenamefont {Peng}, \citenamefont {Chen}, \citenamefont {Chen},
  \citenamefont {Wen}, \citenamefont {Lin}, \citenamefont {Gao},\ and\
  \citenamefont {Liu}}]{Shi2024}%
  \BibitemOpen
  \bibfield  {author} {\bibinfo {author} {\bibfnamefont {A.}~\bibnamefont
  {Shi}}, \bibinfo {author} {\bibfnamefont {Y.}~\bibnamefont {Peng}}, \bibinfo
  {author} {\bibfnamefont {J.}~\bibnamefont {Jiang}}, \bibinfo {author}
  {\bibfnamefont {Y.}~\bibnamefont {Peng}}, \bibinfo {author} {\bibfnamefont
  {P.}~\bibnamefont {Peng}}, \bibinfo {author} {\bibfnamefont {J.}~\bibnamefont
  {Chen}}, \bibinfo {author} {\bibfnamefont {H.}~\bibnamefont {Chen}}, \bibinfo
  {author} {\bibfnamefont {S.}~\bibnamefont {Wen}}, \bibinfo {author}
  {\bibfnamefont {X.}~\bibnamefont {Lin}}, \bibinfo {author} {\bibfnamefont
  {F.}~\bibnamefont {Gao}},\ and\ \bibinfo {author} {\bibfnamefont
  {J.}~\bibnamefont {Liu}},\ }\bibfield  {title} {\bibinfo {title} {Observation
  of topological corner state arrays in photonic quasicrystals},\ }\href
  {https://doi.org/10.1002/lpor.202300956} {\bibfield  {journal} {\bibinfo
  {journal} {Laser \& Photonics Reviews}\ }\textbf {\bibinfo {volume} {18}},\
  \bibinfo {pages} {2300956} (\bibinfo {year} {2024})}\BibitemShut {NoStop}%
\bibitem [{\citenamefont {Vitiello}\ \emph {et~al.}(2014)\citenamefont
  {Vitiello}, \citenamefont {Nobile}, \citenamefont {Ronzani}, \citenamefont
  {Tredicucci}, \citenamefont {Castellano}, \citenamefont {Talora},
  \citenamefont {Li}, \citenamefont {Linfield},\ and\ \citenamefont
  {Davies}}]{Vitiello2014}%
  \BibitemOpen
  \bibfield  {author} {\bibinfo {author} {\bibfnamefont {M.~S.}\ \bibnamefont
  {Vitiello}}, \bibinfo {author} {\bibfnamefont {M.}~\bibnamefont {Nobile}},
  \bibinfo {author} {\bibfnamefont {A.}~\bibnamefont {Ronzani}}, \bibinfo
  {author} {\bibfnamefont {A.}~\bibnamefont {Tredicucci}}, \bibinfo {author}
  {\bibfnamefont {F.}~\bibnamefont {Castellano}}, \bibinfo {author}
  {\bibfnamefont {V.}~\bibnamefont {Talora}}, \bibinfo {author} {\bibfnamefont
  {L.}~\bibnamefont {Li}}, \bibinfo {author} {\bibfnamefont {E.~H.}\
  \bibnamefont {Linfield}},\ and\ \bibinfo {author} {\bibfnamefont {A.~G.}\
  \bibnamefont {Davies}},\ }\bibfield  {title} {\bibinfo {title} {Photonic
  quasi-crystal terahertz lasers},\ }\bibfield  {journal} {\bibinfo  {journal}
  {Nature Communications}\ }\textbf {\bibinfo {volume} {5}},\ \href
  {https://doi.org/10.1038/ncomms6884} {10.1038/ncomms6884} (\bibinfo {year}
  {2014})\BibitemShut {NoStop}%
\bibitem [{\citenamefont {Arjas}\ \emph {et~al.}(2024)\citenamefont {Arjas},
  \citenamefont {Taskinen}, \citenamefont {Heilmann}, \citenamefont {Salerno},\
  and\ \citenamefont {T\"{o}rm\"{a}}}]{Arjas2024}%
  \BibitemOpen
  \bibfield  {author} {\bibinfo {author} {\bibfnamefont {K.}~\bibnamefont
  {Arjas}}, \bibinfo {author} {\bibfnamefont {J.~M.}\ \bibnamefont {Taskinen}},
  \bibinfo {author} {\bibfnamefont {R.}~\bibnamefont {Heilmann}}, \bibinfo
  {author} {\bibfnamefont {G.}~\bibnamefont {Salerno}},\ and\ \bibinfo {author}
  {\bibfnamefont {P.}~\bibnamefont {T\"{o}rm\"{a}}},\ }\bibfield  {title}
  {\bibinfo {title} {High topological charge lasing in quasicrystals},\ }\href
  {https://doi.org/10.1038/s41467-024-53952-5} {\bibfield  {journal} {\bibinfo
  {journal} {Nature Communications}\ }\textbf {\bibinfo {volume} {15}},\
  \bibinfo {pages} {9544} (\bibinfo {year} {2024})}\BibitemShut {NoStop}%
\bibitem [{\citenamefont {Hendrickson}\ \emph {et~al.}(2008)\citenamefont
  {Hendrickson}, \citenamefont {Richards}, \citenamefont {Sweet}, \citenamefont
  {Khitrova}, \citenamefont {Poddubny}, \citenamefont {Ivchenko}, \citenamefont
  {Wegener},\ and\ \citenamefont {Gibbs}}]{Hendrickson2008}%
  \BibitemOpen
  \bibfield  {author} {\bibinfo {author} {\bibfnamefont {J.}~\bibnamefont
  {Hendrickson}}, \bibinfo {author} {\bibfnamefont {B.~C.}\ \bibnamefont
  {Richards}}, \bibinfo {author} {\bibfnamefont {J.}~\bibnamefont {Sweet}},
  \bibinfo {author} {\bibfnamefont {G.}~\bibnamefont {Khitrova}}, \bibinfo
  {author} {\bibfnamefont {A.~N.}\ \bibnamefont {Poddubny}}, \bibinfo {author}
  {\bibfnamefont {E.~L.}\ \bibnamefont {Ivchenko}}, \bibinfo {author}
  {\bibfnamefont {M.}~\bibnamefont {Wegener}},\ and\ \bibinfo {author}
  {\bibfnamefont {H.~M.}\ \bibnamefont {Gibbs}},\ }\bibfield  {title} {\bibinfo
  {title} {Excitonic polaritons in fibonacci quasicrystals},\ }\href
  {https://doi.org/10.1364/oe.16.015382} {\bibfield  {journal} {\bibinfo
  {journal} {Optics Express}\ }\textbf {\bibinfo {volume} {16}},\ \bibinfo
  {pages} {15382} (\bibinfo {year} {2008})}\BibitemShut {NoStop}%
\bibitem [{\citenamefont {Poddubny}\ \emph {et~al.}(2009)\citenamefont
  {Poddubny}, \citenamefont {Pilozzi}, \citenamefont {Voronov},\ and\
  \citenamefont {Ivchenko}}]{Poddubny_PRB2009}%
  \BibitemOpen
  \bibfield  {author} {\bibinfo {author} {\bibfnamefont {A.~N.}\ \bibnamefont
  {Poddubny}}, \bibinfo {author} {\bibfnamefont {L.}~\bibnamefont {Pilozzi}},
  \bibinfo {author} {\bibfnamefont {M.~M.}\ \bibnamefont {Voronov}},\ and\
  \bibinfo {author} {\bibfnamefont {E.~L.}\ \bibnamefont {Ivchenko}},\
  }\bibfield  {title} {\bibinfo {title} {Exciton-polaritonic quasicrystalline
  and aperiodic structures},\ }\href
  {https://doi.org/10.1103/PhysRevB.80.115314} {\bibfield  {journal} {\bibinfo
  {journal} {Phys. Rev. B}\ }\textbf {\bibinfo {volume} {80}},\ \bibinfo
  {pages} {115314} (\bibinfo {year} {2009})}\BibitemShut {NoStop}%
\bibitem [{\citenamefont {Baboux}\ \emph {et~al.}(2017)\citenamefont {Baboux},
  \citenamefont {Levy}, \citenamefont {Lema\^{\i}tre}, \citenamefont {G\'omez},
  \citenamefont {Galopin}, \citenamefont {Le~Gratiet}, \citenamefont {Sagnes},
  \citenamefont {Amo}, \citenamefont {Bloch},\ and\ \citenamefont
  {Akkermans}}]{Baboux_PRB2017}%
  \BibitemOpen
  \bibfield  {author} {\bibinfo {author} {\bibfnamefont {F.}~\bibnamefont
  {Baboux}}, \bibinfo {author} {\bibfnamefont {E.}~\bibnamefont {Levy}},
  \bibinfo {author} {\bibfnamefont {A.}~\bibnamefont {Lema\^{\i}tre}}, \bibinfo
  {author} {\bibfnamefont {C.}~\bibnamefont {G\'omez}}, \bibinfo {author}
  {\bibfnamefont {E.}~\bibnamefont {Galopin}}, \bibinfo {author} {\bibfnamefont
  {L.}~\bibnamefont {Le~Gratiet}}, \bibinfo {author} {\bibfnamefont
  {I.}~\bibnamefont {Sagnes}}, \bibinfo {author} {\bibfnamefont
  {A.}~\bibnamefont {Amo}}, \bibinfo {author} {\bibfnamefont {J.}~\bibnamefont
  {Bloch}},\ and\ \bibinfo {author} {\bibfnamefont {E.}~\bibnamefont
  {Akkermans}},\ }\bibfield  {title} {\bibinfo {title} {Measuring topological
  invariants from generalized edge states in polaritonic quasicrystals},\
  }\href {https://doi.org/10.1103/PhysRevB.95.161114} {\bibfield  {journal}
  {\bibinfo  {journal} {Phys. Rev. B}\ }\textbf {\bibinfo {volume} {95}},\
  \bibinfo {pages} {161114} (\bibinfo {year} {2017})}\BibitemShut {NoStop}%
\bibitem [{\citenamefont {Sturges}\ \emph {et~al.}(2019)\citenamefont
  {Sturges}, \citenamefont {Anderson}, \citenamefont {Buraczewski},
  \citenamefont {Navadeh-Toupchi}, \citenamefont {Adiyatullin}, \citenamefont
  {Jabeen}, \citenamefont {Oberli}, \citenamefont {Portella-Oberli},\ and\
  \citenamefont {Stobińska}}]{Sturges2019}%
  \BibitemOpen
  \bibfield  {author} {\bibinfo {author} {\bibfnamefont {T.~J.}\ \bibnamefont
  {Sturges}}, \bibinfo {author} {\bibfnamefont {M.~D.}\ \bibnamefont
  {Anderson}}, \bibinfo {author} {\bibfnamefont {A.}~\bibnamefont
  {Buraczewski}}, \bibinfo {author} {\bibfnamefont {M.}~\bibnamefont
  {Navadeh-Toupchi}}, \bibinfo {author} {\bibfnamefont {A.~F.}\ \bibnamefont
  {Adiyatullin}}, \bibinfo {author} {\bibfnamefont {F.}~\bibnamefont {Jabeen}},
  \bibinfo {author} {\bibfnamefont {D.~Y.}\ \bibnamefont {Oberli}}, \bibinfo
  {author} {\bibfnamefont {M.~T.}\ \bibnamefont {Portella-Oberli}},\ and\
  \bibinfo {author} {\bibfnamefont {M.}~\bibnamefont {Stobińska}},\ }\bibfield
   {title} {\bibinfo {title} {Anderson localisation in steady states of
  microcavity polaritons},\ }\href {https://doi.org/10.1038/s41598-019-55673-y}
  {\bibfield  {journal} {\bibinfo  {journal} {Scientific Reports}\ }\textbf
  {\bibinfo {volume} {9}},\ \bibinfo {pages} {19396} (\bibinfo {year}
  {2019})}\BibitemShut {NoStop}%
\bibitem [{\citenamefont {Alyatkin}\ \emph {et~al.}(2025)\citenamefont
  {Alyatkin}, \citenamefont {Sitnik}, \citenamefont {Daníelsson},
  \citenamefont {Kartashov}, \citenamefont {T\"{o}pfer}, \citenamefont
  {Sigurðsson},\ and\ \citenamefont {Lagoudakis}}]{Alyatkin2025}%
  \BibitemOpen
  \bibfield  {author} {\bibinfo {author} {\bibfnamefont {S.}~\bibnamefont
  {Alyatkin}}, \bibinfo {author} {\bibfnamefont {K.}~\bibnamefont {Sitnik}},
  \bibinfo {author} {\bibfnamefont {V.~K.}\ \bibnamefont {Daníelsson}},
  \bibinfo {author} {\bibfnamefont {Y.~V.}\ \bibnamefont {Kartashov}}, \bibinfo
  {author} {\bibfnamefont {J.~D.}\ \bibnamefont {T\"{o}pfer}}, \bibinfo
  {author} {\bibfnamefont {H.}~\bibnamefont {Sigurðsson}},\ and\ \bibinfo
  {author} {\bibfnamefont {P.~G.}\ \bibnamefont {Lagoudakis}},\ }\bibfield
  {title} {\bibinfo {title} {Quantum fluids of light in {2D} artificial
  reconfigurable aperiodic crystals with tailored coupling},\ }\href
  {https://doi.org/10.1126/sciadv.adz2484} {\bibfield  {journal} {\bibinfo
  {journal} {Science Advances}\ }\textbf {\bibinfo {volume} {11}},\ \bibinfo
  {pages} {eadz2484} (\bibinfo {year} {2025})}\BibitemShut {NoStop}%
\bibitem [{\citenamefont {Tosi}\ \emph
  {et~al.}(2012{\natexlab{a}})\citenamefont {Tosi}, \citenamefont {Christmann},
  \citenamefont {Berloff}, \citenamefont {Tsotsis}, \citenamefont {Gao},
  \citenamefont {Hatzopoulos}, \citenamefont {Savvidis},\ and\ \citenamefont
  {Baumberg}}]{Tosi2012a}%
  \BibitemOpen
  \bibfield  {author} {\bibinfo {author} {\bibfnamefont {G.}~\bibnamefont
  {Tosi}}, \bibinfo {author} {\bibfnamefont {G.}~\bibnamefont {Christmann}},
  \bibinfo {author} {\bibfnamefont {N.~G.}\ \bibnamefont {Berloff}}, \bibinfo
  {author} {\bibfnamefont {P.}~\bibnamefont {Tsotsis}}, \bibinfo {author}
  {\bibfnamefont {T.}~\bibnamefont {Gao}}, \bibinfo {author} {\bibfnamefont
  {Z.}~\bibnamefont {Hatzopoulos}}, \bibinfo {author} {\bibfnamefont {P.~G.}\
  \bibnamefont {Savvidis}},\ and\ \bibinfo {author} {\bibfnamefont {J.~J.}\
  \bibnamefont {Baumberg}},\ }\bibfield  {title} {\bibinfo {title} {Sculpting
  oscillators with light within a nonlinear quantum fluid},\ }\href
  {https://doi.org/10.1038/nphys2182} {\bibfield  {journal} {\bibinfo
  {journal} {Nature Physics}\ }\textbf {\bibinfo {volume} {8}},\ \bibinfo
  {pages} {190–194} (\bibinfo {year} {2012}{\natexlab{a}})}\BibitemShut
  {NoStop}%
\bibitem [{\citenamefont {Ohadi}\ \emph {et~al.}(2016)\citenamefont {Ohadi},
  \citenamefont {Gregory}, \citenamefont {Freegarde}, \citenamefont {Rubo},
  \citenamefont {Kavokin}, \citenamefont {Berloff},\ and\ \citenamefont
  {Lagoudakis}}]{Ohadi_PRX2016}%
  \BibitemOpen
  \bibfield  {author} {\bibinfo {author} {\bibfnamefont {H.}~\bibnamefont
  {Ohadi}}, \bibinfo {author} {\bibfnamefont {R.~L.}\ \bibnamefont {Gregory}},
  \bibinfo {author} {\bibfnamefont {T.}~\bibnamefont {Freegarde}}, \bibinfo
  {author} {\bibfnamefont {Y.~G.}\ \bibnamefont {Rubo}}, \bibinfo {author}
  {\bibfnamefont {A.~V.}\ \bibnamefont {Kavokin}}, \bibinfo {author}
  {\bibfnamefont {N.~G.}\ \bibnamefont {Berloff}},\ and\ \bibinfo {author}
  {\bibfnamefont {P.~G.}\ \bibnamefont {Lagoudakis}},\ }\bibfield  {title}
  {\bibinfo {title} {Nontrivial phase coupling in polariton multiplets},\
  }\href {https://doi.org/10.1103/PhysRevX.6.031032} {\bibfield  {journal}
  {\bibinfo  {journal} {Phys. Rev. X}\ }\textbf {\bibinfo {volume} {6}},\
  \bibinfo {pages} {031032} (\bibinfo {year} {2016})}\BibitemShut {NoStop}%
\bibitem [{\citenamefont {T\"{o}pfer}\ \emph {et~al.}(2020)\citenamefont
  {T\"{o}pfer}, \citenamefont {Sigurdsson}, \citenamefont {Pickup},\ and\
  \citenamefont {Lagoudakis}}]{Topfer2020}%
  \BibitemOpen
  \bibfield  {author} {\bibinfo {author} {\bibfnamefont {J.~D.}\ \bibnamefont
  {T\"{o}pfer}}, \bibinfo {author} {\bibfnamefont {H.}~\bibnamefont
  {Sigurdsson}}, \bibinfo {author} {\bibfnamefont {L.}~\bibnamefont {Pickup}},\
  and\ \bibinfo {author} {\bibfnamefont {P.~G.}\ \bibnamefont {Lagoudakis}},\
  }\bibfield  {title} {\bibinfo {title} {Time-delay polaritonics},\ }\bibfield
  {journal} {\bibinfo  {journal} {Communications Physics}\ }\textbf {\bibinfo
  {volume} {3}},\ \href {https://doi.org/10.1038/s42005-019-0271-0}
  {10.1038/s42005-019-0271-0} (\bibinfo {year} {2020})\BibitemShut {NoStop}%
\bibitem [{\citenamefont {Deng}\ \emph {et~al.}(2010)\citenamefont {Deng},
  \citenamefont {Haug},\ and\ \citenamefont {Yamamoto}}]{Deng_RMP2010}%
  \BibitemOpen
  \bibfield  {author} {\bibinfo {author} {\bibfnamefont {H.}~\bibnamefont
  {Deng}}, \bibinfo {author} {\bibfnamefont {H.}~\bibnamefont {Haug}},\ and\
  \bibinfo {author} {\bibfnamefont {Y.}~\bibnamefont {Yamamoto}},\ }\bibfield
  {title} {\bibinfo {title} {Exciton-polariton bose-einstein condensation},\
  }\href {https://doi.org/10.1103/RevModPhys.82.1489} {\bibfield  {journal}
  {\bibinfo  {journal} {Rev. Mod. Phys.}\ }\textbf {\bibinfo {volume} {82}},\
  \bibinfo {pages} {1489} (\bibinfo {year} {2010})}\BibitemShut {NoStop}%
\bibitem [{\citenamefont {Boulier}\ \emph {et~al.}(2020)\citenamefont
  {Boulier}, \citenamefont {Jacquet}, \citenamefont {Maître}, \citenamefont
  {Lerario}, \citenamefont {Claude}, \citenamefont {Pigeon}, \citenamefont
  {Glorieux}, \citenamefont {Amo}, \citenamefont {Bloch}, \citenamefont
  {Bramati},\ and\ \citenamefont {Giacobino}}]{Boulier2020}%
  \BibitemOpen
  \bibfield  {author} {\bibinfo {author} {\bibfnamefont {T.}~\bibnamefont
  {Boulier}}, \bibinfo {author} {\bibfnamefont {M.~J.}\ \bibnamefont
  {Jacquet}}, \bibinfo {author} {\bibfnamefont {A.}~\bibnamefont {Maître}},
  \bibinfo {author} {\bibfnamefont {G.}~\bibnamefont {Lerario}}, \bibinfo
  {author} {\bibfnamefont {F.}~\bibnamefont {Claude}}, \bibinfo {author}
  {\bibfnamefont {S.}~\bibnamefont {Pigeon}}, \bibinfo {author} {\bibfnamefont
  {Q.}~\bibnamefont {Glorieux}}, \bibinfo {author} {\bibfnamefont
  {A.}~\bibnamefont {Amo}}, \bibinfo {author} {\bibfnamefont {J.}~\bibnamefont
  {Bloch}}, \bibinfo {author} {\bibfnamefont {A.}~\bibnamefont {Bramati}},\
  and\ \bibinfo {author} {\bibfnamefont {E.}~\bibnamefont {Giacobino}},\
  }\bibfield  {title} {\bibinfo {title} {Microcavity polaritons for quantum
  simulation},\ }\bibfield  {journal} {\bibinfo  {journal} {Advanced Quantum
  Technologies}\ }\textbf {\bibinfo {volume} {3}},\ \href
  {https://doi.org/10.1002/qute.202000052} {10.1002/qute.202000052} (\bibinfo
  {year} {2020})\BibitemShut {NoStop}%
\bibitem [{\citenamefont {Berloff}\ \emph {et~al.}(2017)\citenamefont
  {Berloff}, \citenamefont {Silva}, \citenamefont {Kalinin}, \citenamefont
  {Askitopoulos}, \citenamefont {T\"{o}pfer}, \citenamefont {Cilibrizzi},
  \citenamefont {Langbein},\ and\ \citenamefont {Lagoudakis}}]{Berloff2017}%
  \BibitemOpen
  \bibfield  {author} {\bibinfo {author} {\bibfnamefont {N.~G.}\ \bibnamefont
  {Berloff}}, \bibinfo {author} {\bibfnamefont {M.}~\bibnamefont {Silva}},
  \bibinfo {author} {\bibfnamefont {K.}~\bibnamefont {Kalinin}}, \bibinfo
  {author} {\bibfnamefont {A.}~\bibnamefont {Askitopoulos}}, \bibinfo {author}
  {\bibfnamefont {J.~D.}\ \bibnamefont {T\"{o}pfer}}, \bibinfo {author}
  {\bibfnamefont {P.}~\bibnamefont {Cilibrizzi}}, \bibinfo {author}
  {\bibfnamefont {W.}~\bibnamefont {Langbein}},\ and\ \bibinfo {author}
  {\bibfnamefont {P.~G.}\ \bibnamefont {Lagoudakis}},\ }\bibfield  {title}
  {\bibinfo {title} {Realizing the classical {XY} hamiltonian in polariton
  simulators},\ }\href {https://doi.org/10.1038/nmat4971} {\bibfield  {journal}
  {\bibinfo  {journal} {Nature Materials}\ }\textbf {\bibinfo {volume} {16}},\
  \bibinfo {pages} {1120–1126} (\bibinfo {year} {2017})}\BibitemShut
  {NoStop}%
\bibitem [{\citenamefont {Alyatkin}\ \emph {et~al.}(2021)\citenamefont
  {Alyatkin}, \citenamefont {Sigurdsson}, \citenamefont {Askitopoulos},
  \citenamefont {T\"{o}pfer},\ and\ \citenamefont {Lagoudakis}}]{Alyatkin2021}%
  \BibitemOpen
  \bibfield  {author} {\bibinfo {author} {\bibfnamefont {S.}~\bibnamefont
  {Alyatkin}}, \bibinfo {author} {\bibfnamefont {H.}~\bibnamefont
  {Sigurdsson}}, \bibinfo {author} {\bibfnamefont {A.}~\bibnamefont
  {Askitopoulos}}, \bibinfo {author} {\bibfnamefont {J.~D.}\ \bibnamefont
  {T\"{o}pfer}},\ and\ \bibinfo {author} {\bibfnamefont {P.~G.}\ \bibnamefont
  {Lagoudakis}},\ }\bibfield  {title} {\bibinfo {title} {Quantum fluids of
  light in all-optical scatterer lattices},\ }\href
  {https://doi.org/10.1038/s41467-021-25845-4} {\bibfield  {journal} {\bibinfo
  {journal} {Nature Communications}\ }\textbf {\bibinfo {volume} {12}},\
  \bibinfo {pages} {5571} (\bibinfo {year} {2021})}\BibitemShut {NoStop}%
\bibitem [{\citenamefont {Alyatkin}\ \emph {et~al.}(2024)\citenamefont
  {Alyatkin}, \citenamefont {Sigurðsson}, \citenamefont {Kartashov},
  \citenamefont {Gnusov}, \citenamefont {Sitnik}, \citenamefont {T\"{o}pfer},\
  and\ \citenamefont {Lagoudakis}}]{Alyatkin2024}%
  \BibitemOpen
  \bibfield  {author} {\bibinfo {author} {\bibfnamefont {S.}~\bibnamefont
  {Alyatkin}}, \bibinfo {author} {\bibfnamefont {H.}~\bibnamefont
  {Sigurðsson}}, \bibinfo {author} {\bibfnamefont {Y.~V.}\ \bibnamefont
  {Kartashov}}, \bibinfo {author} {\bibfnamefont {I.}~\bibnamefont {Gnusov}},
  \bibinfo {author} {\bibfnamefont {K.}~\bibnamefont {Sitnik}}, \bibinfo
  {author} {\bibfnamefont {J.~D.}\ \bibnamefont {T\"{o}pfer}},\ and\ \bibinfo
  {author} {\bibfnamefont {P.~G.}\ \bibnamefont {Lagoudakis}},\ }\bibfield
  {title} {\bibinfo {title} {All-optical triangular and honeycomb lattices of
  exciton–polaritons},\ }\href {https://doi.org/10.1063/5.0180272} {\bibfield
   {journal} {\bibinfo  {journal} {Applied Physics Letters}\ }\textbf {\bibinfo
  {volume} {124}},\ \bibinfo {pages} {062105} (\bibinfo {year}
  {2024})}\BibitemShut {NoStop}%
\bibitem [{\citenamefont {Alyatkin}\ \emph {et~al.}(2026)\citenamefont
  {Alyatkin}, \citenamefont {Kartashov}, \citenamefont {Sitnik}, \citenamefont
  {Grigoryev},\ and\ \citenamefont {Lagoudakis}}]{alyatkin2026}%
  \BibitemOpen
  \bibfield  {author} {\bibinfo {author} {\bibfnamefont {S.}~\bibnamefont
  {Alyatkin}}, \bibinfo {author} {\bibfnamefont {Y.~V.}\ \bibnamefont
  {Kartashov}}, \bibinfo {author} {\bibfnamefont {K.}~\bibnamefont {Sitnik}},
  \bibinfo {author} {\bibfnamefont {P.}~\bibnamefont {Grigoryev}},\ and\
  \bibinfo {author} {\bibfnamefont {P.~G.}\ \bibnamefont {Lagoudakis}},\ }\href
  {https://arxiv.org/abs/2605.13206} {\bibinfo {title} {Observation of an
  aperiodic polariton monotile}} (\bibinfo {year} {2026}),\ \Eprint
  {https://arxiv.org/abs/2605.13206} {arXiv:2605.13206 [cond-mat.quant-gas]}
  \BibitemShut {NoStop}%
\bibitem [{\citenamefont {Smith}\ \emph
  {et~al.}(2024{\natexlab{a}})\citenamefont {Smith}, \citenamefont {Myers},
  \citenamefont {Kaplan},\ and\ \citenamefont {Goodman-Strauss}}]{Smith2024}%
  \BibitemOpen
  \bibfield  {author} {\bibinfo {author} {\bibfnamefont {D.}~\bibnamefont
  {Smith}}, \bibinfo {author} {\bibfnamefont {J.~S.}\ \bibnamefont {Myers}},
  \bibinfo {author} {\bibfnamefont {C.~S.}\ \bibnamefont {Kaplan}},\ and\
  \bibinfo {author} {\bibfnamefont {C.}~\bibnamefont {Goodman-Strauss}},\
  }\bibfield  {title} {\bibinfo {title} {An aperiodic monotile},\ }\bibfield
  {journal} {\bibinfo  {journal} {Combinatorial Theory}\ }\textbf {\bibinfo
  {volume} {4}},\ \href {https://doi.org/10.5070/c64163843} {10.5070/c64163843}
  (\bibinfo {year} {2024}{\natexlab{a}})\BibitemShut {NoStop}%
\bibitem [{\citenamefont {Socolar}(2023)}]{Socolar_PRB2023}%
  \BibitemOpen
  \bibfield  {author} {\bibinfo {author} {\bibfnamefont {J.~E.~S.}\
  \bibnamefont {Socolar}},\ }\bibfield  {title} {\bibinfo {title}
  {Quasicrystalline structure of the hat monotile tilings},\ }\href
  {https://doi.org/10.1103/PhysRevB.108.224109} {\bibfield  {journal} {\bibinfo
   {journal} {Phys. Rev. B}\ }\textbf {\bibinfo {volume} {108}},\ \bibinfo
  {pages} {224109} (\bibinfo {year} {2023})}\BibitemShut {NoStop}%
\bibitem [{cod()}]{code}%
  \BibitemOpen
  \href@noop {} {}\bibinfo {note}
  {\url{https://github.com/fixgoats/basichatmonotiler}}\BibitemShut {NoStop}%
\bibitem [{\citenamefont {Goblot}\ \emph {et~al.}(2020)\citenamefont {Goblot},
  \citenamefont {Štrkalj}, \citenamefont {Pernet}, \citenamefont {Lado},
  \citenamefont {Dorow}, \citenamefont {Lemaître}, \citenamefont {Le~Gratiet},
  \citenamefont {Harouri}, \citenamefont {Sagnes}, \citenamefont {Ravets},
  \citenamefont {Amo}, \citenamefont {Bloch},\ and\ \citenamefont
  {Zilberberg}}]{Goblot2020}%
  \BibitemOpen
  \bibfield  {author} {\bibinfo {author} {\bibfnamefont {V.}~\bibnamefont
  {Goblot}}, \bibinfo {author} {\bibfnamefont {A.}~\bibnamefont {Štrkalj}},
  \bibinfo {author} {\bibfnamefont {N.}~\bibnamefont {Pernet}}, \bibinfo
  {author} {\bibfnamefont {J.~L.}\ \bibnamefont {Lado}}, \bibinfo {author}
  {\bibfnamefont {C.}~\bibnamefont {Dorow}}, \bibinfo {author} {\bibfnamefont
  {A.}~\bibnamefont {Lemaître}}, \bibinfo {author} {\bibfnamefont
  {L.}~\bibnamefont {Le~Gratiet}}, \bibinfo {author} {\bibfnamefont
  {A.}~\bibnamefont {Harouri}}, \bibinfo {author} {\bibfnamefont
  {I.}~\bibnamefont {Sagnes}}, \bibinfo {author} {\bibfnamefont
  {S.}~\bibnamefont {Ravets}}, \bibinfo {author} {\bibfnamefont
  {A.}~\bibnamefont {Amo}}, \bibinfo {author} {\bibfnamefont {J.}~\bibnamefont
  {Bloch}},\ and\ \bibinfo {author} {\bibfnamefont {O.}~\bibnamefont
  {Zilberberg}},\ }\bibfield  {title} {\bibinfo {title} {Emergence of
  criticality through a cascade of delocalization transitions in quasiperiodic
  chains},\ }\href {https://doi.org/10.1038/s41567-020-0908-7} {\bibfield
  {journal} {\bibinfo  {journal} {Nature Physics}\ }\textbf {\bibinfo {volume}
  {16}},\ \bibinfo {pages} {832–836} (\bibinfo {year} {2020})}\BibitemShut
  {NoStop}%
\bibitem [{\citenamefont {Evers}\ and\ \citenamefont
  {Mirlin}(2008)}]{Evers_RevModPhys2008}%
  \BibitemOpen
  \bibfield  {author} {\bibinfo {author} {\bibfnamefont {F.}~\bibnamefont
  {Evers}}\ and\ \bibinfo {author} {\bibfnamefont {A.~D.}\ \bibnamefont
  {Mirlin}},\ }\bibfield  {title} {\bibinfo {title} {Anderson transitions},\
  }\href {https://doi.org/10.1103/RevModPhys.80.1355} {\bibfield  {journal}
  {\bibinfo  {journal} {Rev. Mod. Phys.}\ }\textbf {\bibinfo {volume} {80}},\
  \bibinfo {pages} {1355} (\bibinfo {year} {2008})}\BibitemShut {NoStop}%
\bibitem [{\citenamefont {Saul}\ \emph {et~al.}(1988)\citenamefont {Saul},
  \citenamefont {Llois},\ and\ \citenamefont {Weissmann}}]{Saul1988}%
  \BibitemOpen
  \bibfield  {author} {\bibinfo {author} {\bibfnamefont {A.}~\bibnamefont
  {Saul}}, \bibinfo {author} {\bibfnamefont {A.~M.}\ \bibnamefont {Llois}},\
  and\ \bibinfo {author} {\bibfnamefont {M.}~\bibnamefont {Weissmann}},\
  }\bibfield  {title} {\bibinfo {title} {Wavefunctions of one-dimensional
  incommensurate hamiltonians: the fractal dimension and its relationship with
  localisation},\ }\href {https://doi.org/10.1088/0022-3719/21/11/006}
  {\bibfield  {journal} {\bibinfo  {journal} {Journal of Physics C: Solid State
  Physics}\ }\textbf {\bibinfo {volume} {21}},\ \bibinfo {pages} {2137–2151}
  (\bibinfo {year} {1988})}\BibitemShut {NoStop}%
\bibitem [{\citenamefont {Mac\'e}\ \emph {et~al.}(2017)\citenamefont {Mac\'e},
  \citenamefont {Jagannathan}, \citenamefont {Kalugin}, \citenamefont
  {Mosseri},\ and\ \citenamefont {Pi\'echon}}]{Mace_PRB2017}%
  \BibitemOpen
  \bibfield  {author} {\bibinfo {author} {\bibfnamefont {N.}~\bibnamefont
  {Mac\'e}}, \bibinfo {author} {\bibfnamefont {A.}~\bibnamefont {Jagannathan}},
  \bibinfo {author} {\bibfnamefont {P.}~\bibnamefont {Kalugin}}, \bibinfo
  {author} {\bibfnamefont {R.}~\bibnamefont {Mosseri}},\ and\ \bibinfo {author}
  {\bibfnamefont {F.}~\bibnamefont {Pi\'echon}},\ }\bibfield  {title} {\bibinfo
  {title} {Critical eigenstates and their properties in one- and
  two-dimensional quasicrystals},\ }\href
  {https://doi.org/10.1103/PhysRevB.96.045138} {\bibfield  {journal} {\bibinfo
  {journal} {Phys. Rev. B}\ }\textbf {\bibinfo {volume} {96}},\ \bibinfo
  {pages} {045138} (\bibinfo {year} {2017})}\BibitemShut {NoStop}%
\bibitem [{\citenamefont {Zhu}\ \emph {et~al.}(2024)\citenamefont {Zhu},
  \citenamefont {Yu}, \citenamefont {Johnstone},\ and\ \citenamefont
  {Sanchez-Palencia}}]{Zhu_PRA2024}%
  \BibitemOpen
  \bibfield  {author} {\bibinfo {author} {\bibfnamefont {Z.}~\bibnamefont
  {Zhu}}, \bibinfo {author} {\bibfnamefont {S.}~\bibnamefont {Yu}}, \bibinfo
  {author} {\bibfnamefont {D.}~\bibnamefont {Johnstone}},\ and\ \bibinfo
  {author} {\bibfnamefont {L.}~\bibnamefont {Sanchez-Palencia}},\ }\bibfield
  {title} {\bibinfo {title} {Localization and spectral structure in
  two-dimensional quasicrystal potentials},\ }\href
  {https://doi.org/10.1103/PhysRevA.109.013314} {\bibfield  {journal} {\bibinfo
   {journal} {Phys. Rev. A}\ }\textbf {\bibinfo {volume} {109}},\ \bibinfo
  {pages} {013314} (\bibinfo {year} {2024})}\BibitemShut {NoStop}%
\bibitem [{\citenamefont {Duncan}(2024)}]{Duncan_PRB2024}%
  \BibitemOpen
  \bibfield  {author} {\bibinfo {author} {\bibfnamefont {C.~W.}\ \bibnamefont
  {Duncan}},\ }\bibfield  {title} {\bibinfo {title} {Critical states and
  anomalous mobility edges in two-dimensional diagonal quasicrystals},\ }\href
  {https://doi.org/10.1103/PhysRevB.109.014210} {\bibfield  {journal} {\bibinfo
   {journal} {Phys. Rev. B}\ }\textbf {\bibinfo {volume} {109}},\ \bibinfo
  {pages} {014210} (\bibinfo {year} {2024})}\BibitemShut {NoStop}%
\bibitem [{\citenamefont {Koga}(2024)}]{Koga2024}%
  \BibitemOpen
  \bibfield  {author} {\bibinfo {author} {\bibfnamefont {A.}~\bibnamefont
  {Koga}},\ }\bibfield  {title} {\bibinfo {title} {Multifractality and
  hyperuniformity in quasicrystals},\ }\bibfield  {journal} {\bibinfo
  {journal} {JPSJ News and Comments}\ }\textbf {\bibinfo {volume} {21}},\ \href
  {https://doi.org/10.7566/jpsjnc.21.21} {10.7566/jpsjnc.21.21} (\bibinfo
  {year} {2024})\BibitemShut {NoStop}%
\bibitem [{\citenamefont {Lima}\ \emph {et~al.}(2005)\citenamefont {Lima},
  \citenamefont {de~Moura}, \citenamefont {Lyra},\ and\ \citenamefont
  {Nazareno}}]{Lima_2005PRB}%
  \BibitemOpen
  \bibfield  {author} {\bibinfo {author} {\bibfnamefont {R.~P.~A.}\
  \bibnamefont {Lima}}, \bibinfo {author} {\bibfnamefont {F.~A. B.~F.}\
  \bibnamefont {de~Moura}}, \bibinfo {author} {\bibfnamefont {M.~L.}\
  \bibnamefont {Lyra}},\ and\ \bibinfo {author} {\bibfnamefont {H.~N.}\
  \bibnamefont {Nazareno}},\ }\bibfield  {title} {\bibinfo {title} {Critical
  wave-packet dynamics in the power-law bond disordered anderson model},\
  }\href {https://doi.org/10.1103/PhysRevB.71.235112} {\bibfield  {journal}
  {\bibinfo  {journal} {Phys. Rev. B}\ }\textbf {\bibinfo {volume} {71}},\
  \bibinfo {pages} {235112} (\bibinfo {year} {2005})}\BibitemShut {NoStop}%
\bibitem [{\citenamefont {Yuan}\ \emph {et~al.}(2000)\citenamefont {Yuan},
  \citenamefont {Grimm}, \citenamefont {Repetowicz},\ and\ \citenamefont
  {Schreiber}}]{Yuan_PRB2000}%
  \BibitemOpen
  \bibfield  {author} {\bibinfo {author} {\bibfnamefont {H.~Q.}\ \bibnamefont
  {Yuan}}, \bibinfo {author} {\bibfnamefont {U.}~\bibnamefont {Grimm}},
  \bibinfo {author} {\bibfnamefont {P.}~\bibnamefont {Repetowicz}},\ and\
  \bibinfo {author} {\bibfnamefont {M.}~\bibnamefont {Schreiber}},\ }\bibfield
  {title} {\bibinfo {title} {Energy spectra, wave functions, and quantum
  diffusion for quasiperiodic systems},\ }\href
  {https://doi.org/10.1103/PhysRevB.62.15569} {\bibfield  {journal} {\bibinfo
  {journal} {Phys. Rev. B}\ }\textbf {\bibinfo {volume} {62}},\ \bibinfo
  {pages} {15569} (\bibinfo {year} {2000})}\BibitemShut {NoStop}%
\bibitem [{\citenamefont {Segev}\ \emph {et~al.}(2013)\citenamefont {Segev},
  \citenamefont {Silberberg},\ and\ \citenamefont
  {Christodoulides}}]{Segev2013}%
  \BibitemOpen
  \bibfield  {author} {\bibinfo {author} {\bibfnamefont {M.}~\bibnamefont
  {Segev}}, \bibinfo {author} {\bibfnamefont {Y.}~\bibnamefont {Silberberg}},\
  and\ \bibinfo {author} {\bibfnamefont {D.~N.}\ \bibnamefont
  {Christodoulides}},\ }\bibfield  {title} {\bibinfo {title} {Anderson
  localization of light},\ }\href {https://doi.org/10.1038/nphoton.2013.30}
  {\bibfield  {journal} {\bibinfo  {journal} {Nature Photonics}\ }\textbf
  {\bibinfo {volume} {7}},\ \bibinfo {pages} {197–204} (\bibinfo {year}
  {2013})}\BibitemShut {NoStop}%
\bibitem [{\citenamefont {Ketzmerick}\ \emph {et~al.}(1997)\citenamefont
  {Ketzmerick}, \citenamefont {Kruse}, \citenamefont {Kraut},\ and\
  \citenamefont {Geisel}}]{Ketzmerick_PRL1997}%
  \BibitemOpen
  \bibfield  {author} {\bibinfo {author} {\bibfnamefont {R.}~\bibnamefont
  {Ketzmerick}}, \bibinfo {author} {\bibfnamefont {K.}~\bibnamefont {Kruse}},
  \bibinfo {author} {\bibfnamefont {S.}~\bibnamefont {Kraut}},\ and\ \bibinfo
  {author} {\bibfnamefont {T.}~\bibnamefont {Geisel}},\ }\bibfield  {title}
  {\bibinfo {title} {What determines the spreading of a wave packet?},\ }\href
  {https://doi.org/10.1103/PhysRevLett.79.1959} {\bibfield  {journal} {\bibinfo
   {journal} {Phys. Rev. Lett.}\ }\textbf {\bibinfo {volume} {79}},\ \bibinfo
  {pages} {1959} (\bibinfo {year} {1997})}\BibitemShut {NoStop}%
\bibitem [{\citenamefont {Zhong}\ and\ \citenamefont
  {Mosseri}(1995)}]{Zhong1995}%
  \BibitemOpen
  \bibfield  {author} {\bibinfo {author} {\bibfnamefont {J.~X.}\ \bibnamefont
  {Zhong}}\ and\ \bibinfo {author} {\bibfnamefont {R.}~\bibnamefont
  {Mosseri}},\ }\bibfield  {title} {\bibinfo {title} {Quantum dynamics in
  quasiperiodic systems},\ }\href {https://doi.org/10.1088/0953-8984/7/44/008}
  {\bibfield  {journal} {\bibinfo  {journal} {Journal of Physics: Condensed
  Matter}\ }\textbf {\bibinfo {volume} {7}},\ \bibinfo {pages} {8383–8404}
  (\bibinfo {year} {1995})}\BibitemShut {NoStop}%
\bibitem [{\citenamefont {Ketzmerick}\ \emph {et~al.}(1992)\citenamefont
  {Ketzmerick}, \citenamefont {Petschel},\ and\ \citenamefont
  {Geisel}}]{Ketzmerick_PRL1992}%
  \BibitemOpen
  \bibfield  {author} {\bibinfo {author} {\bibfnamefont {R.}~\bibnamefont
  {Ketzmerick}}, \bibinfo {author} {\bibfnamefont {G.}~\bibnamefont
  {Petschel}},\ and\ \bibinfo {author} {\bibfnamefont {T.}~\bibnamefont
  {Geisel}},\ }\bibfield  {title} {\bibinfo {title} {Slow decay of temporal
  correlations in quantum systems with cantor spectra},\ }\href
  {https://doi.org/10.1103/PhysRevLett.69.695} {\bibfield  {journal} {\bibinfo
  {journal} {Phys. Rev. Lett.}\ }\textbf {\bibinfo {volume} {69}},\ \bibinfo
  {pages} {695} (\bibinfo {year} {1992})}\BibitemShut {NoStop}%
\bibitem [{\citenamefont {Eichelkraut}\ \emph {et~al.}(2013)\citenamefont
  {Eichelkraut}, \citenamefont {Heilmann}, \citenamefont {Weimann},
  \citenamefont {St\"{u}tzer}, \citenamefont {Dreisow}, \citenamefont
  {Christodoulides}, \citenamefont {Nolte},\ and\ \citenamefont
  {Szameit}}]{Eichelkraut2013}%
  \BibitemOpen
  \bibfield  {author} {\bibinfo {author} {\bibfnamefont {T.}~\bibnamefont
  {Eichelkraut}}, \bibinfo {author} {\bibfnamefont {R.}~\bibnamefont
  {Heilmann}}, \bibinfo {author} {\bibfnamefont {S.}~\bibnamefont {Weimann}},
  \bibinfo {author} {\bibfnamefont {S.}~\bibnamefont {St\"{u}tzer}}, \bibinfo
  {author} {\bibfnamefont {F.}~\bibnamefont {Dreisow}}, \bibinfo {author}
  {\bibfnamefont {D.~N.}\ \bibnamefont {Christodoulides}}, \bibinfo {author}
  {\bibfnamefont {S.}~\bibnamefont {Nolte}},\ and\ \bibinfo {author}
  {\bibfnamefont {A.}~\bibnamefont {Szameit}},\ }\bibfield  {title} {\bibinfo
  {title} {Mobility transition from ballistic to diffusive transport in
  non-hermitian lattices},\ }\href {https://doi.org/10.1038/ncomms3533}
  {\bibfield  {journal} {\bibinfo  {journal} {Nature Communications}\ }\textbf
  {\bibinfo {volume} {4}},\ \bibinfo {pages} {2533} (\bibinfo {year}
  {2013})}\BibitemShut {NoStop}%
\bibitem [{\citenamefont {Richard}\ \emph {et~al.}(2005)\citenamefont
  {Richard}, \citenamefont {Kasprzak}, \citenamefont {Romestain}, \citenamefont
  {Andr\'e},\ and\ \citenamefont {Dang}}]{Richard_PRL2005}%
  \BibitemOpen
  \bibfield  {author} {\bibinfo {author} {\bibfnamefont {M.}~\bibnamefont
  {Richard}}, \bibinfo {author} {\bibfnamefont {J.}~\bibnamefont {Kasprzak}},
  \bibinfo {author} {\bibfnamefont {R.}~\bibnamefont {Romestain}}, \bibinfo
  {author} {\bibfnamefont {R.}~\bibnamefont {Andr\'e}},\ and\ \bibinfo {author}
  {\bibfnamefont {L.~S.}\ \bibnamefont {Dang}},\ }\bibfield  {title} {\bibinfo
  {title} {Spontaneous coherent phase transition of polaritons in cdte
  microcavities},\ }\href {https://doi.org/10.1103/PhysRevLett.94.187401}
  {\bibfield  {journal} {\bibinfo  {journal} {Phys. Rev. Lett.}\ }\textbf
  {\bibinfo {volume} {94}},\ \bibinfo {pages} {187401} (\bibinfo {year}
  {2005})}\BibitemShut {NoStop}%
\bibitem [{\citenamefont {Alyatkin}\ \emph {et~al.}(2020)\citenamefont
  {Alyatkin}, \citenamefont {T\"opfer}, \citenamefont {Askitopoulos},
  \citenamefont {Sigurdsson},\ and\ \citenamefont {Lagoudakis}}]{Alyatkin2020}%
  \BibitemOpen
  \bibfield  {author} {\bibinfo {author} {\bibfnamefont {S.}~\bibnamefont
  {Alyatkin}}, \bibinfo {author} {\bibfnamefont {J.~D.}\ \bibnamefont
  {T\"opfer}}, \bibinfo {author} {\bibfnamefont {A.}~\bibnamefont
  {Askitopoulos}}, \bibinfo {author} {\bibfnamefont {H.}~\bibnamefont
  {Sigurdsson}},\ and\ \bibinfo {author} {\bibfnamefont {P.~G.}\ \bibnamefont
  {Lagoudakis}},\ }\bibfield  {title} {\bibinfo {title} {Optical control of
  couplings in polariton condensate lattices},\ }\href
  {https://doi.org/10.1103/PhysRevLett.124.207402} {\bibfield  {journal}
  {\bibinfo  {journal} {Phys. Rev. Lett.}\ }\textbf {\bibinfo {volume} {124}},\
  \bibinfo {pages} {207402} (\bibinfo {year} {2020})}\BibitemShut {NoStop}%
\bibitem [{\citenamefont {Tosi}\ \emph
  {et~al.}(2012{\natexlab{b}})\citenamefont {Tosi}, \citenamefont {Christmann},
  \citenamefont {Berloff}, \citenamefont {Tsotsis}, \citenamefont {Gao},
  \citenamefont {Hatzopoulos}, \citenamefont {Savvidis},\ and\ \citenamefont
  {Baumberg}}]{Tosi2012b}%
  \BibitemOpen
  \bibfield  {author} {\bibinfo {author} {\bibfnamefont {G.}~\bibnamefont
  {Tosi}}, \bibinfo {author} {\bibfnamefont {G.}~\bibnamefont {Christmann}},
  \bibinfo {author} {\bibfnamefont {N.}~\bibnamefont {Berloff}}, \bibinfo
  {author} {\bibfnamefont {P.}~\bibnamefont {Tsotsis}}, \bibinfo {author}
  {\bibfnamefont {T.}~\bibnamefont {Gao}}, \bibinfo {author} {\bibfnamefont
  {Z.}~\bibnamefont {Hatzopoulos}}, \bibinfo {author} {\bibfnamefont
  {P.}~\bibnamefont {Savvidis}},\ and\ \bibinfo {author} {\bibfnamefont
  {J.}~\bibnamefont {Baumberg}},\ }\bibfield  {title} {\bibinfo {title}
  {Geometrically locked vortex lattices in semiconductor quantum fluids},\
  }\bibfield  {journal} {\bibinfo  {journal} {Nature Communications}\ }\textbf
  {\bibinfo {volume} {3}},\ \href {https://doi.org/10.1038/ncomms2255}
  {10.1038/ncomms2255} (\bibinfo {year} {2012}{\natexlab{b}})\BibitemShut
  {NoStop}%
\bibitem [{\citenamefont {Wouters}\ \emph {et~al.}(2008)\citenamefont
  {Wouters}, \citenamefont {Carusotto},\ and\ \citenamefont
  {Ciuti}}]{Wouters_PRB2008}%
  \BibitemOpen
  \bibfield  {author} {\bibinfo {author} {\bibfnamefont {M.}~\bibnamefont
  {Wouters}}, \bibinfo {author} {\bibfnamefont {I.}~\bibnamefont {Carusotto}},\
  and\ \bibinfo {author} {\bibfnamefont {C.}~\bibnamefont {Ciuti}},\ }\bibfield
   {title} {\bibinfo {title} {Spatial and spectral shape of inhomogeneous
  nonequilibrium exciton-polariton condensates},\ }\href
  {https://doi.org/10.1103/PhysRevB.77.115340} {\bibfield  {journal} {\bibinfo
  {journal} {Phys. Rev. B}\ }\textbf {\bibinfo {volume} {77}},\ \bibinfo
  {pages} {115340} (\bibinfo {year} {2008})}\BibitemShut {NoStop}%
\bibitem [{\citenamefont {Zhong}\ \emph {et~al.}(2001)\citenamefont {Zhong},
  \citenamefont {Diener}, \citenamefont {Steck}, \citenamefont {Oskay},
  \citenamefont {Raizen}, \citenamefont {Plummer}, \citenamefont {Zhang},\ and\
  \citenamefont {Niu}}]{Zhong_PRL2001}%
  \BibitemOpen
  \bibfield  {author} {\bibinfo {author} {\bibfnamefont {J.}~\bibnamefont
  {Zhong}}, \bibinfo {author} {\bibfnamefont {R.~B.}\ \bibnamefont {Diener}},
  \bibinfo {author} {\bibfnamefont {D.~A.}\ \bibnamefont {Steck}}, \bibinfo
  {author} {\bibfnamefont {W.~H.}\ \bibnamefont {Oskay}}, \bibinfo {author}
  {\bibfnamefont {M.~G.}\ \bibnamefont {Raizen}}, \bibinfo {author}
  {\bibfnamefont {E.~W.}\ \bibnamefont {Plummer}}, \bibinfo {author}
  {\bibfnamefont {Z.}~\bibnamefont {Zhang}},\ and\ \bibinfo {author}
  {\bibfnamefont {Q.}~\bibnamefont {Niu}},\ }\bibfield  {title} {\bibinfo
  {title} {Shape of the quantum diffusion front},\ }\href
  {https://doi.org/10.1103/PhysRevLett.86.2485} {\bibfield  {journal} {\bibinfo
   {journal} {Phys. Rev. Lett.}\ }\textbf {\bibinfo {volume} {86}},\ \bibinfo
  {pages} {2485} (\bibinfo {year} {2001})}\BibitemShut {NoStop}%
\bibitem [{\citenamefont {del Valle Inclan~Redondo}\ \emph
  {et~al.}(2024)\citenamefont {del Valle Inclan~Redondo}, \citenamefont {Xu},
  \citenamefont {Liew}, \citenamefont {Ostrovskaya}, \citenamefont {Stegmaier},
  \citenamefont {Thomale}, \citenamefont {Schneider}, \citenamefont {Dam},
  \citenamefont {Klembt}, \citenamefont {H\"{o}fling}, \citenamefont
  {Tarucha},\ and\ \citenamefont {Fraser}}]{delValleInclanRedondo2024}%
  \BibitemOpen
  \bibfield  {author} {\bibinfo {author} {\bibfnamefont {Y.}~\bibnamefont {del
  Valle Inclan~Redondo}}, \bibinfo {author} {\bibfnamefont {X.}~\bibnamefont
  {Xu}}, \bibinfo {author} {\bibfnamefont {T.~C.~H.}\ \bibnamefont {Liew}},
  \bibinfo {author} {\bibfnamefont {E.~A.}\ \bibnamefont {Ostrovskaya}},
  \bibinfo {author} {\bibfnamefont {A.}~\bibnamefont {Stegmaier}}, \bibinfo
  {author} {\bibfnamefont {R.}~\bibnamefont {Thomale}}, \bibinfo {author}
  {\bibfnamefont {C.}~\bibnamefont {Schneider}}, \bibinfo {author}
  {\bibfnamefont {S.}~\bibnamefont {Dam}}, \bibinfo {author} {\bibfnamefont
  {S.}~\bibnamefont {Klembt}}, \bibinfo {author} {\bibfnamefont
  {S.}~\bibnamefont {H\"{o}fling}}, \bibinfo {author} {\bibfnamefont
  {S.}~\bibnamefont {Tarucha}},\ and\ \bibinfo {author} {\bibfnamefont {M.~D.}\
  \bibnamefont {Fraser}},\ }\bibfield  {title} {\bibinfo {title}
  {Non-reciprocal band structures in an exciton–polariton floquet optical
  lattice},\ }\href {https://doi.org/10.1038/s41566-024-01424-z} {\bibfield
  {journal} {\bibinfo  {journal} {Nature Photonics}\ }\textbf {\bibinfo
  {volume} {18}},\ \bibinfo {pages} {548–553} (\bibinfo {year}
  {2024})}\BibitemShut {NoStop}%
\bibitem [{\citenamefont {Schirmann}\ \emph {et~al.}(2024)\citenamefont
  {Schirmann}, \citenamefont {Franca}, \citenamefont {Flicker},\ and\
  \citenamefont {Grushin}}]{Schirmann_PRL2024}%
  \BibitemOpen
  \bibfield  {author} {\bibinfo {author} {\bibfnamefont {J.}~\bibnamefont
  {Schirmann}}, \bibinfo {author} {\bibfnamefont {S.}~\bibnamefont {Franca}},
  \bibinfo {author} {\bibfnamefont {F.}~\bibnamefont {Flicker}},\ and\ \bibinfo
  {author} {\bibfnamefont {A.~G.}\ \bibnamefont {Grushin}},\ }\bibfield
  {title} {\bibinfo {title} {Physical properties of an aperiodic monotile with
  graphene-like features, chirality, and zero modes},\ }\href
  {https://doi.org/10.1103/PhysRevLett.132.086402} {\bibfield  {journal}
  {\bibinfo  {journal} {Phys. Rev. Lett.}\ }\textbf {\bibinfo {volume} {132}},\
  \bibinfo {pages} {086402} (\bibinfo {year} {2024})}\BibitemShut {NoStop}%
\bibitem [{\citenamefont {Smith}\ \emph
  {et~al.}(2024{\natexlab{b}})\citenamefont {Smith}, \citenamefont {Myers},
  \citenamefont {Kaplan},\ and\ \citenamefont {Goodman-Strauss}}]{Smith2024_b}%
  \BibitemOpen
  \bibfield  {author} {\bibinfo {author} {\bibfnamefont {D.}~\bibnamefont
  {Smith}}, \bibinfo {author} {\bibfnamefont {J.~S.}\ \bibnamefont {Myers}},
  \bibinfo {author} {\bibfnamefont {C.~S.}\ \bibnamefont {Kaplan}},\ and\
  \bibinfo {author} {\bibfnamefont {C.}~\bibnamefont {Goodman-Strauss}},\
  }\bibfield  {title} {\bibinfo {title} {A chiral aperiodic monotile},\
  }\bibfield  {journal} {\bibinfo  {journal} {Combinatorial Theory}\ }\textbf
  {\bibinfo {volume} {4}},\ \href {https://doi.org/10.5070/c64264241}
  {10.5070/c64264241} (\bibinfo {year} {2024}{\natexlab{b}})\BibitemShut
  {NoStop}%
\end{thebibliography}
\end{document}